\documentclass[reprint,amsmath,amssymb,aps,prd,nofootinbib]{revtex4-1}

\usepackage{amssymb,amsmath,amsfonts}
\usepackage{mathrsfs}
\usepackage{dsfont} 
\usepackage{graphicx,bm}
\usepackage{multirow}
\usepackage{color}
\usepackage{enumerate}
\usepackage{placeins}
\usepackage{dcolumn}
\usepackage{array}
\usepackage{bbold} 
\usepackage{lipsum}  

\usepackage{subfigure}
\usepackage{subfig}

\newcommand{\Tr}{\mathrm{Tr}}
\newcommand{\be}{\begin{equation}}
\newcommand{\ee}{\end{equation}}
\newcommand{\bea}{\begin{align}}
\newcommand{\eea}{\end{align}}
\newcommand{\Det}{\mathrm{Det}}  
\newcommand{\diag}{\mathop{\mathrm{diag}}}

\newcommand{\matt}{\text{matt}}
\newcommand{\Pol}{\text{Pol}}
\newcommand{\glue}{\text{glue}}
\newcommand{\conf}{\text{conf}}
\newcommand{\GeV}{\text{GeV}}

\providecommand{\e}[1]{\ensuremath{{\scriptscriptstyle E\negthinspace #1}}}

\allowdisplaybreaks

\begin{document}

\title{The fate of the critical endpoint at large $N_c$}


\author{P{\'e}ter Kov{\'a}cs}
\email{kovacs.peter@wigner.hu}
\affiliation{
Institute for Particle and Nuclear Physics, Wigner Research Centre for Physics, 1121 Budapest, Hungary
}
\affiliation{
Institute of Physics, E\"otv\"os University, 1117 Budapest, Hungary
}
\author{Gy\H{o}z\H{o} Kov{\'a}cs}
\affiliation{
Institute for Particle and Nuclear Physics, Wigner Research Centre for Physics, 1121 Budapest, Hungary
}
\affiliation{
Institute of Physics, E\"otv\"os University, 1117 Budapest, Hungary
}

\author{Francesco Giacosa}
\affiliation{Institute of Physics, Jan Kochanowski University, ul. Uniwersytecka 7, P-25-406 Kielce, Poland}
\affiliation{Institute for Theoretical Physics, Goethe-University, Max-von-Laue-Str. 1, D-60438 Frankfurt am Main, Germany}

\begin{abstract}

The phase diagram of QCD is investigated by varying number of colors $N_c$ 
within a Polyakov loop quark-meson chiral model. In particular, our attention is focused on the critical point(s): the critical point present for $N_c=3$ moves toward the $\mu_q$-axis and disappears as soon as the number of color is increased. Yet, a distinct critical point emerges along the temperature axis for $N_c = 53$ and moves toward finite density when increasing $N_c$ further. Thus, the phase diagram at large $N_c$ looks specular w.r.t. the $N_c = 3$ results, with the first order transition in the upper-left and crossover in the down-right regions of the the $(\mu_q,T)$-plane. The pressure is also evaluated in dependence of $N_c$, showing a scaling with $N_c^0$ in the confined and chirally broken phase and with $N_c^2$ in the deconfined one. Moreover, the presence of a chirally symmetric but confined `quarkyonic phase' at large density and moderate temperature with a pressure proportional to $N_c$ is confirmed.
\end{abstract}

\maketitle

\section{Introduction}


The phase-diagram of QCD is one of the main subjects of high energy physics and is in the centre of numerous theoretical, numerical, and experimental works \cite{Rischke:2003mt,Ratti}. 

Various experimental collaborations that focus on heavy ion collisions were and are able to investigate certain areas of the quark chemical potential-temperature $(\mu_q,T)$ plane ($\mu_B=3\mu_q$ is the baryonic chemical potential), in dependence of the energy and the types of nuclei involved in the collisions \cite{Florkowski:2010zz}. 

At the same time, lattice QCD numerical simulations achieved a great precision along the $T$-axis at $\mu_q=0$ \cite{Borsanyi:2012ve,Karsch:2003jg,Bazavov:2019www} and are constantly improving along the positive $\mu_q$ axis, which is notoriously complicated by the fermionic sign problem \cite{Langelage:2009jb,Vovchenko:2018zgt}.
                                    
Moreover, numerous models based on (global) symmetries of QCD and involving mesons or quarks degrees of freedom (d.o.f.)---or eventually both of them---delivered a consistent qualitative picture of the main features of the diagram that fits well with both experimental and lattice outcomes, even though they still differ in quantitative details \cite{Dumitru:2000in,Scavenius:2002ru,Kovacs:2016juc,Hansen:2006ee,Folkestad:2018psc,Nishimura:2018wla,Nishimura:2017kjl,Dumitru:2012fw,Lo:2013hla}. 

A crossover phase transition along the $T$-axis and a first order one along the $\mu_q$-axis are expected. In between, a critical end point (CEP), whose precise location is the main subject of numerous studies, with a second-order transition emerges. According to various approaches, the so-called confinement/deconfinement phase transitions and the chiral phase transition seem to coincide---or lie very close to each other---on the $\mu_q$-$T$ plane. Of course, this is the picture for the number of color $N_c=3$ realized in Nature. 

Another interesting approach to QCD is the so-called large-$N_c$ limit \cite{tHooft:1974pnl,Witten:1979kh,Hollowood:2013ax}. Namely, in this limit certain simplifications take place: the quark-antiquark mesons (as well as glueballs) become stable, because the interaction between them vanishes when $N_c \rightarrow \infty$ (the exact scaling behaviors shall be described later on). Quite remarkably, $N_c=3$ can be seen as a ``large number'' in some cases, since the implications of the large-$N_c$ approach are well confirmed; for instance, the $J/\psi$ (as well as other charmonia below the $\bar{D}D$ threshold) is very narrow, and the decay of the---mainly strange-antistrange---$f_2^{\prime}(1525)$ meson into two pions is extremely small, even though the phase space is large. 

The natural question concerns the properties of the phase diagram in the large-$N_c$ limit. Does it keep some of the $N_c=3$ features or is it completely different? Do the chiral and confinement/deconfinement phase transitions coincide? 

In previous works on this topic, it was indeed found that the phase diagram for $N_c=3$ is rather different from its large-$N_c$ counterpart. According to Refs. \cite{McLerran:2007qj,McLerran:2008ua} a quarkyonic phase, still confined but chirally symmetric, is expected to take place at high quark chemical potential and moderate temperatures. Later on, this view was confirmed in various works. The fate of nuclear matter was also discussed in the large-$N_c$ limit, the outcome being that it might be rather a fortunate outcome of our $N_c=3$ world \cite{Bonanno:2011yr} (thus, $N_c=3$ should not be regarded as a ``large number'' for the binding of nuclei). Different aspects of large-$N_c$ have been studied in a variety of works, see Refs.
 \cite{Cohen:2020tgr,Giacosa:2017mis,Hidaka:2011jj,McLerran:2009nr,Toublan:2005rq,PhysRevD.29.1222,Heinz:2011xq,Panero:2009tv,Lucini:2012gg,Lottini:2011zp} and refs. therein.
 
Here, we intend to use a chiral model for QCD, the so-called Polyakov-loop extended linear sigma model (PLeLSM), in order to study the phase diagram at large-$N_c$.
The model is based on both (pseudo)scalar and (axial-)vector chiral multiplets and has been investigated in the vacuum in Refs. \cite{Parganlija:2010fz,Parganlija:2012fy}. Later on, it has been applied to the QCD medium \cite{Kovacs:2014cua,Kovacs:2016juc,Kovacs:2021ger} by coupling it to quarks as well as the to the Polyakov loop that describes, in a thermodynamic sense, confinement. 

The extension of the PLeLSM to large $N_c$ is straightforward for what concerns mesons and quarks, but care is needed for the Polyakov-loop sector \cite{Fukushima:2017csk}. In the main text we use the the so-called uniform eigenvalue Ansatz, while, for comparison, in Appendix \ref{App:phi_n_phi_n} we employ  the $\Phi_n = \Phi^n$ approximation. Both deliver qualitatively similar results.

The outcomes turn out to be quite interesting: we do confirm the existence of a confined and chirally symmetric region at intermediate densities and low temperatures (a quarkyonic phase), which in turn implies that the chiral and the deconfinement phase transitions do not coincide in the large-$N_c$ limit. 

Moreover, we also observe other remarkable phenomena:
(i) The CEP disappears very fast when increasing $N_c$, since already for  $N_c=4$ it is not present. It means that the CEP with a crossover line on its left and a first-order one on its right is solely a feature of the $N_c=3$ world. At large-$N_c$, the $\mu_q$-axis features a crossover transition.
(ii) For intermediate $N_c$ (in the range $(4,52)$) the whole diagram contains only cross-over transitions, (iii)
A new CEP along the $T$-axis emerges when $N_c$ is large enough ($N_c=53$). This CEP then moves toward larger and larger $\mu_B$ for increasing $N_c$. This is in agreement with the gluons dominating matter along the $T$-axis with a first order confinement/deconfinement transition, e.g. Refs. \cite{Drago:2001gd,Kondo:2015noa,Lacroix:2012pt,Pisarski:1983db} 

In summary, the large-$N_c$ phase diagram is utterly different from the $N_c=3$ one, thus showing that $N_c=3$ is in this respect `not large' with the original CEP being a property of $N_c=3$ only. Yet, the quarkyonic phase is confirmed in the large-$N_c$ domain and can be interpreted as one of the features that link $N_c=3$ to the large-$N_c$ limit.

The paper is organized as follows. In Sec. \ref{sec:model_intro} and \ref{sec:large_Nc} we briefly introduce the model, summarize the basic properties of the large-$N_c$ approach and give the explicit $N_c$ dependence of the parameters of our Lagrangian. We also show the $N_c$ scaling of the meson condensates and the meson masses in the vacuum. In Sec.~\ref{sec:polyakov_Nc} we discuss the Polyakov-loops at arbitrary $N_c$, establish a Polyakov-loop potential suitable for the $N_c>3$ calculations and compatible with the so-called uniform eigenvalue Ansatz. The corresponding grand potential and the field equations are also presented here. In Sec.~\ref{sec:results} we show our results for the $N_c$ dependence of the phase diagram together with the $N_c$ scaling of the pressure as a function of temperature. Finally, we conclude in Sec.~\ref{sec:conclusion}.

\section{The PLeLSM model}
\label{sec:model_intro}

The model that we use is a three flavored vector and axial-vector extended linear sigma model with Polyakov-loop and quark variables. At zero temperature, a version of this model was investigated thoroughly in \cite{Parganlija:2012fy}, while at finite temperature in \cite{Kovacs:2016juc}.  The Lagrangian consists of a mesonic and a Yukawa part, 
\begin{equation}
    \mathcal{L} = \mathcal{L}_{m} + \mathcal{L}_{Y}
    \text{ ,}
\end{equation}
while the Polyakov-loop is introduced with the help of the grand potential (see later). The meson sector contains four nonets, namely the pseudoscalar $P$, the scalar $S$, the vector $V$ and the axial vector $A$ matrix fields,
\be \begin{split} \label{Eq:mfields}
M=S+i P=\sum_a \left(S_a + iP_a\right)T_a \text{ ,}\\
L^\mu =V^\mu + A^\mu = \sum_a \left(V^\mu_a +A^\mu_a \right)T_a \text{ ,} \\
R^\mu =V^\mu - A^\mu = \sum_a \left(V^\mu_a -A^\mu_a \right)T_a \text{ ,} \\
\end{split} \ee
where $T_a (a=0\dots 8)$ denotes the generators of $U(3)$. The mesonic part of the Lagrangian reads
\begin{widetext}
\be
\begin{split}
\label{Eq:lagm}
\mathcal{L}_{m} =&\Tr \left[ \left( D_\mu M \right)^\dagger \left( D^\mu M \right) \right] - m_0 \Tr \left( M^\dagger M \right) - \lambda_1 \left[ \Tr \left( M^\dagger M \right) \right]^2  - \lambda_2 \left[ \Tr \left( M^\dagger M \right)^2 \right] \\
&+ c \left( \det M + \det M^\dagger \right) + \Tr \left[ H \left( M + M^\dagger \right) \right]  - \frac{1}{4} \Tr \left[ L_{\mu\nu}L^{\mu\nu}+R_{\mu\nu}R^{\mu\nu} \right] \\
&+\Tr \left[\left(\frac{m_1^2 }{2} +\Delta \right) \left( L_\mu L^\mu + R_\mu R^\mu \right) \right]  +\frac{h_1}{2} \Tr \left( \phi^\dagger \phi \right) \Tr \left[ L_\mu L^\mu +R_\mu R^\mu \right] \\
&+ h_2 \Tr \left[ \left( M R_\mu \right)^\dagger \left( M R^\mu \right) + \left( L_\mu  M \right)^\dagger \left( L^\mu M  \right) \right]  + 2 h_3 \Tr \left[ R_\mu M^\dagger L^\mu M \right] \\ 
& -2 g_2  \Tr \lbrace L_{\mu\nu} \left[ L^\mu , L^\nu \right] \rbrace + \Tr \lbrace R_{\mu\nu} \left[ R^\mu , R^\nu \right]  ,
\end{split}
\ee
where
\be 
\begin{split}
D^\mu &=\partial^\mu M - i  g_1(L_\mu M -M R_\mu) -ie A^\mu \left[T_3,M \right], \\
L^{\mu\nu}&=\partial^\mu L^\nu -ie A^\mu \left[ T_3,L^\nu\right] -\lbrace \partial^\nu L^\mu -ie A^\nu \left[ T_3,L^\mu\right] \rbrace, \\
R^{\mu\nu}&=\partial^\mu R^\nu -ie A^\mu \left[ T_3,R^\nu\right] -\lbrace \partial^\nu R^\mu -ie A^\nu \left[ T_3,R^\mu\right] \rbrace
,\\
\end{split} 
\ee 
\end{widetext}
and the explicit symmetry breaking terms are
\be \begin{split}
H&=H_0 T_0 + H_8 T_8 = \frac{1}{2} \diag \left( h_{N},h_{N},\sqrt{2}h_{S} \right),\\
\Delta&=\Delta_0 T_0 + \Delta_8 T_8 =  \diag \left( \delta_{N},\delta_{N},\delta_{S} \right) \text{ .}
\end{split} \ee
The mesonic Lagrangian contains the dynamical and the meson-meson interaction terms up to fourth order, that are chirally symmetric ($SU(3)_L\times SU(3)_R \times U(1)_V\times U(1)_A$). 
Explicit symmetry breaking terms (proportional to $H =         \diag (h_N,h_N,h_S)$ and $\Delta = \diag (\delta_N,\delta_N,\delta_S)$) and $U(1)_A$ anomaly term (proportional to $c$) are also included. 

In the fermionic sector of the model, $N_f=2+1$ constituent quarks are present in a Yukawa-type Lagrangian 
\be
\begin{split}
\label{Eq:lagY}
\mathcal{L}_{Y} =&\bar\psi \left( i \gamma_\mu \partial^\mu -g_F (S+i \gamma_5 P)\right) \psi .
\end{split}
\ee
It should be noted here that constituent quarks could also be coupled to the (axial-)vector nonets, as it is discussed e.g. in \cite{Kovacs:2021kas}. However, they are not relevant for the aim of the present study and are thus omitted, 

The model parameters are the bare masses $m_0^2$ and $m_1^2$, the couplings $g_1$, $g_2$, $\lambda_1$, $\lambda_2$, $h_1$, $h_2$ and $h_3$, the already mentioned symmetry breaking external fields $h_{N/S}$ and  $\delta_{N/S}$, the $U_A(1)$ anomaly parameter $c$, and, finally, the fermion-meson coupling $g_F$. 
These parameters are determined with a $\chi^2$ minimization method using tree-level meson masses and decay widths as physical inputs. The tree-level masses and decay widths can be calculated after applying the spontaneous symmetry breaking (SSB) and shifting the corresponding fields with their nonzero vacuum expectation values. Here we assume two scalar condensates, the $\phi_{N/S}\equiv \langle \sigma_{N/S}\rangle$ nonstrange and strange condensates.  More details about the model and the fitting procedure can be found in \cite{Kovacs:2016juc}. For completeness, the parameter values found in \cite{Kovacs:2016juc} are listed in Table~\ref{Tab:param} as set A. 
\begin{table}[!htb]
  \caption{Parameter sets.  Left column is taken from \cite{Kovacs:2016juc} (set A) and right column is taken from \cite{Kovacs:2021ger} (set B) \label{Tab:param}}
\centering
\begin{tabular}[c]{c c c}\hline\hline
Parameter & Set A  & Set B \\\hline
$\phi_{N}$ [GeV]    & $0.1411$       & $0.1290$\\
$\phi_{S}$ [GeV]    & $0.1416$       & $0.1406$\\
$m_{0}^2$ [GeV$^2$] & $2.3925\e{-4}$ & $-1.2370\e{-2}$\\
$m_{1}^2$ [GeV$^2$] & $6.3298\e{-8}$ & $0.5600$\\
$\lambda_{1}$       & $-1.6738$      & $-1.0096$\\
$\lambda_{2}$       & $23.5078$      & $25.7328$\\
$c_{1}$ [GeV]       & $1.3086$       & $1.4700$\\
$\delta_{S}$ [GeV$^2$] & $0.1133$    & $0.2305$\\

$g_{1}$ & $5.6156$        & $5.3295$\\
$g_{2}$ & $3.0467$       & $-1.0579$\\
$h_{1}$ & $37.4617$       & $5.8467$\\
$h_{2}$ & $4.2281$       & $-12.3456$\\
$h_{3}$ & $2.9839$       &  $3.5755$\\
$g_{F}$ & $4.5708$       & $4.9571$\\
$M_0$ [GeV]  & $0.3511$        & $0.3935$\\
\hline\hline
\end{tabular}
\end{table}
It is worth to note that in \cite{Kovacs:2021ger} the same model with additional nonzero vector condensates was applied for investigation of the properties of compact stars. Through the investigation of the asymptotic behavior of the $\phi_{N/S}$ nonstrange and strange scalar condensates it was found that the following condition
\be 
\label{Eq:NS_cond}
\frac{3}{2}h_1+h_2+h_3<0
\ee
is needed for the condensates to vanish---as expected in the chirally symmetric phase---for very large values of $\mu_B$. This restriction also has an advantage at large $N_c$, which will be explained later. Consequently, we have taken another parameter set from \cite{Kovacs:2021ger} (set B in Table~\ref{Tab:param}), which complies this requirement.

\subsection{Grand potential and field equations}
\label{ssec:omega_FE}

As it is discussed in detail in \cite{Kovacs:2016juc}, the thermodynamic behavior of the system can be determined by the calculation of the grand potential $\Omega (T,\mu_q)$, which---in the so-called hybrid approximation---consist of a tree-level mesonic part, a one-loop level fermionic part with vanishing mesonic fields, and a Polyakov-loop potential:
\be 
\Omega (T,\mu_q )= U(\langle M\rangle ) + \Omega_{\bar qq}^{(0)} (T,\mu_q) + U(\langle \Phi \rangle,\langle \bar \Phi \rangle) \text{ .}
\label{Eq:grand_pot}
\ee 
In that approximation we assumed altogether four order parameters, the $\phi_{N/S}$ nonstrange and strange scalar condensates and the $\Phi$ and $\bar \Phi$ Polyakov-loop variables. The field equations are given by the saddle point of the grand potential with respect to the four order parameters 
\be 
\label{Eq:field_eq}
\frac{\partial \Omega (T,\mu_q )}{\partial \phi_N}=\frac{\partial \Omega (T,\mu_q )}{\partial \phi_S}=\frac{\partial \Omega (T,\mu_q )}{\partial \Phi }=\frac{\partial \Omega (T,\mu_q )}{\partial \bar \Phi}=0,
\ee 
which can be solved for non-zero $T$ and/or $\mu_q$ using the parameters fixed at $T=\mu_q=0$. It is worth to note that at zero temperature $\Phi = \bar \Phi \equiv 0$, thus there are only two field equations. From the solution of these two equations at $T=\mu_q=0$ the $h_{N/S}$ external fields can be determined. 

\section{Large $N_c$}
\label{sec:large_Nc}

\subsection{Main properties of large $N_c$}

As argued by G. 't Hooft \cite{tHooft:1973alw,tHooft:1974pnl}, the coupling parameter of QCD, denoted as $g_{QCD}$,is not a free parameter in the sense that it takes part in the setting of the QCD scale. Since, beside the current quark masses, this is the only parameter of the QCD Lagrangian, we have no evident expansion parameter, which would be required for a systematic expansion. One possible solution is to enlarge the $SU(N_c)$ gauge group of the theory---from $N_c=3$ to $N_c>3$---and use $1/N_c$ as an expansion parameter. As it turns out, QCD substantially simplifies at the leading order of the $1/N_c$ expansion, under the assumption that $N_c g_{QCD}^2$ is kept fixed as $N_c\to \infty$. Basically, the properties of the large-$N_c$ approximation arise from the combinatorial factors of the various Feynman diagrams for large number of colors.
The main properties are (see also Refs. \cite{Witten:1979kh,Lebed:1998st}):
\begin{itemize}
    \item The $\bar{q}q$ mesons and glueballs are free, non-interacting and stable particles.
    \item For the $\bar{q}q$ meson, each decay  amplitude runs (at most) with $1/\sqrt{N_c}$, while each four-leg  scattering amplitudes (at most) with $1/N_c$.
    \item The leading contributions to the elastic scattering amplitude are given by tree-level graphs with mesons as mediating particles. This result can be related to the Regge phenomenology. 
    \item Diagrams that falls apart by cutting an internal gluon line are large-$N_c$ suppressed. This is the so called Zweig or Okubo-Zweig-Iizuka (OZI) rule.
    \item The baryon masses diverge with 
    $\sim N_c$, but the quark masses are $N_c$ independent.
\end{itemize}

It is also known (e.g., \cite{McLerran:2007qj}) that the pseudocritical temperature of the chiral phase transition at $\mu_B=0$ is independent of $N_c$ ($T_c\propto N_c^0$). 
Since in the hadronic phase the relevant degrees of freedom are color-singlet $\bar{q}q$ mesons and glueballs, while in the quark gluon plasma (QGP) phase they are colored quarks ($\propto N_c$) and, predominantly, gluons ($\propto N_c^2$), one can argue that the pressure $p$ and the energy density $\epsilon$ scale as $\propto N_c^0$ for small and $\propto N_c^2$ for large temperatures \cite{McLerran:2007qj, Cohen:2020tgr}.

\subsection{$N_c$ dependence of the model parameters}
\label{ssec:param_Nc_dep}

According to \cite{Witten:1979kh} and \cite{Parganlija:2010fz}, the large-$N_c$ scaling of the parameters of the PLeLSM Lagrangian are summarized in Table~\ref{tab:Nc_dep}. 
\begin{table}[h!]
    \centering
    \begin{tabular}{|c|c|}
    \hline
        $m_0^2,\ m_1^2,\ \delta_S$ & $N_c^0$ \\
        $g_1,\ g_2,\ g_f$ & $1/\sqrt{N_c}$\\
        $\lambda_2,\ h_2,\ h_3$ & $N_c^{-1}$ \\
        $\lambda_1,\ h_1$ & $N_c^{-2}$ \\
        $c_1$ & $N_c^{-3/2}$ \\
        $h_{N/S}$ & $\sqrt{N_c}$\\
        $g_F$ & $1/\sqrt{N_c}$\\
         \hline
    \end{tabular}
    \caption{$N_c$ dependence of the parameters}
    \label{tab:Nc_dep}
\end{table}
In more detail, the $k$-leg meson vertex scales as $\Gamma_k \propto N_c^{1-\frac{k}{2}}$, consequently the $g_1$ and $g_2$ parameters---being three-leg-couplings---scale as $N_c^{-1/2}$,  while the $\lambda_2$, $h_2$ and $h_3$ parameters---being four-leg couplings---scale as $N_c^{-1}$. 
Due to the different trace structure (square of trace of two meson fields instead of trace of four meson fields), the parameters $\lambda_1$ and $h_1$ are more suppressed and scale with $N_c^{-2}$. The parameters $m_0^2$, $m_1^2$, $\delta_S$ correspond to tree-level meson mass terms, hence they are independent of $N_c$. Since the $U(1)_A$ anomaly has an extra $1/N_c$ suppression \cite{Witten:1979vv}, $c_1$ scales as $N_c^{-3/2}$.
From the Goldstone-theorem and PCAC relation it can be deduced that the $h_{N/S}$ external fields scale as $\sqrt{N_c}$. It should be noted that from the PCAC we also expect that the $\phi_{N/S}$ meson condensates scale similarly, i.e. $\sqrt{N_c}$, however, we only set the scaling of $h_{N/S}$ (since they are connected through the field equations and cannot be scaled separately). 

\begin{figure}[htbp]
    \centering
    \includegraphics[width=0.48\textwidth]{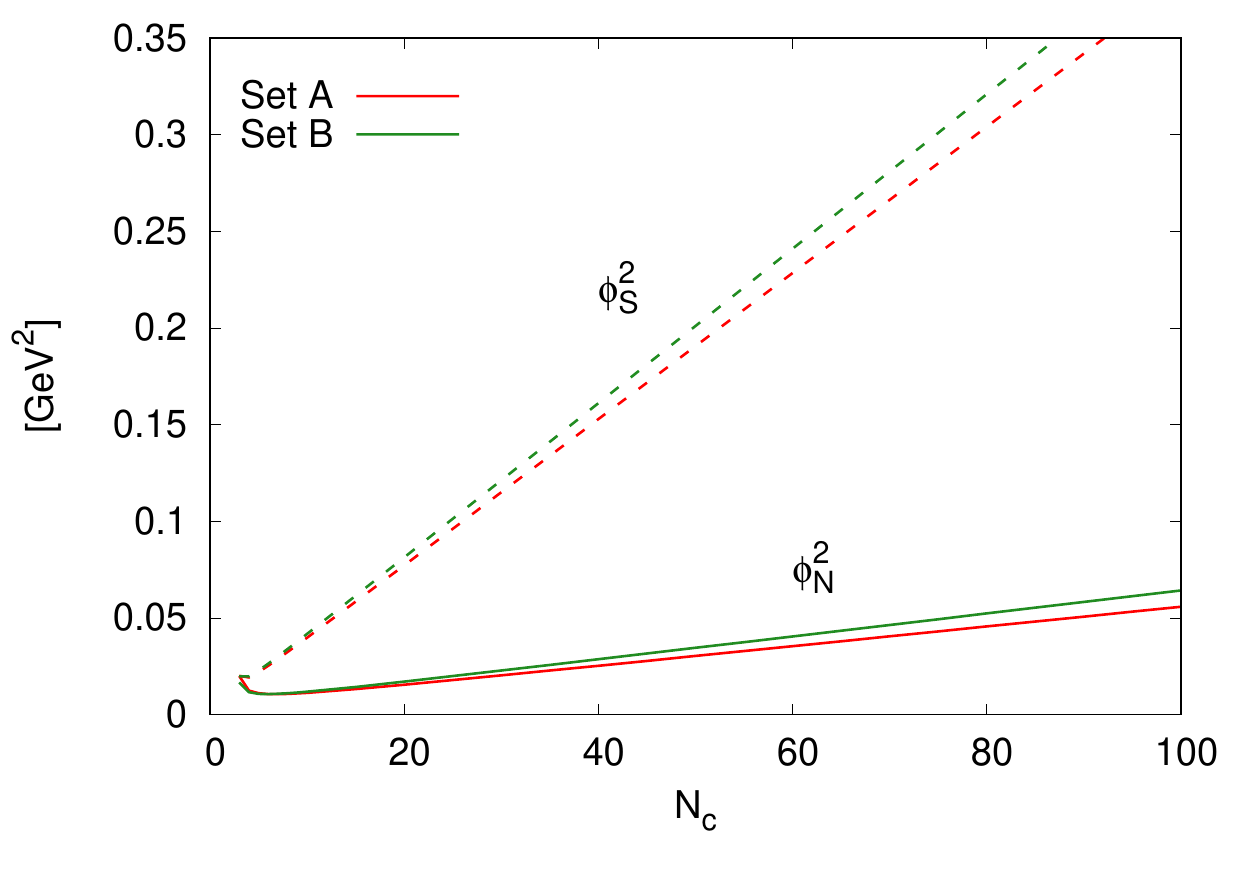} 
    \caption{$N_c$ dependence of the squared meson condensates $\phi_{N}^2$ and $\phi_S^2$ with using the parameter set A, listed in the left column (red) and set B, listed in the right column (green) of Tab.~\ref{Tab:param}}
    \label{fig:phi_Nc}
\end{figure}

In practice, we implement the large $N_c$ dependence as follows:
\begin{equation}
    g_1(N_c) = \sqrt{\frac{3}{N_c}}g_1(N_c=3)
\end{equation}
and similarly for all the other parameters in Table~\ref{tab:Nc_dep}. 
These rescalings are done at $T=\mu_q=0$, than the coupled filed equations are solved for non-zero $T/\mu_q$. It is worth to note that the $h_N$, $h_S$ external fields are calculated at $T=\mu_q=0$ for $N_c=3$ from the field equations and rescaled to a desired $N_c$. We then solve the field equations again\footnote{Note, an appropriate choice of the initial values is extremely important to find a physically meaningful solution -- i.e. global minima of the grand potential that is in connection to the solution at $N_c=3$ -- for the field equations.} to get the values of $\phi_{N/S}$ condensates for the new $N_c$.
In Fig.~\ref{fig:phi_Nc} the $N_c$ dependence of the $\phi_{N/S}$ condensates is shown using both parameter sets in Table~\ref{Tab:param}.
As it can be seen, after a rather abrupt change for small $N_c$ values, the condensates show the expected scaling behavior (linear for the squared condensates). 

\begin{figure*}[htbp]
        \centering
            \includegraphics[width=0.98\textwidth]{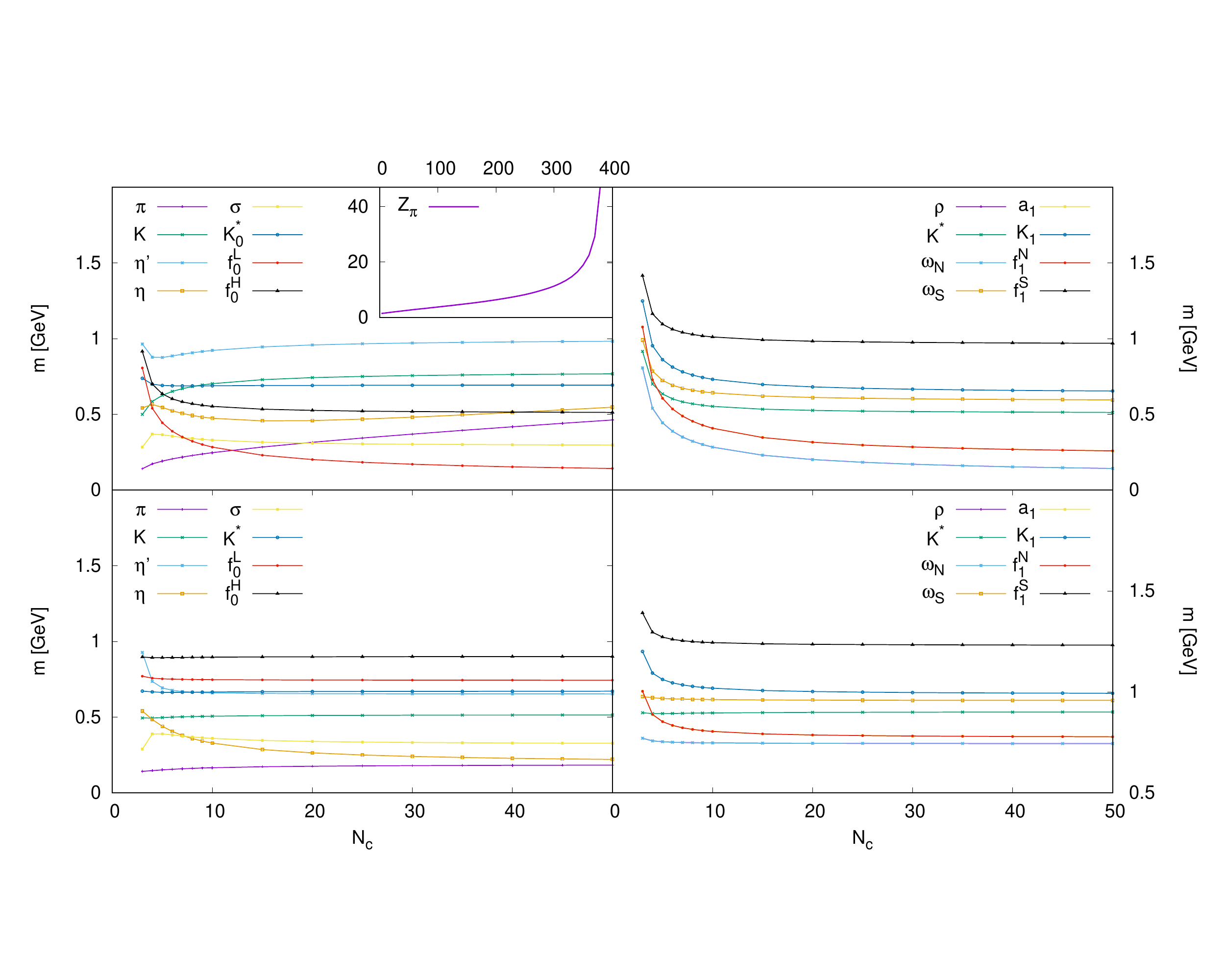}  
        \caption{The $N_c$ dependence of the scalar and pseudoscalar (left) and the vector and axial-vector (right) meson masses. In the top figures the parameter set from \cite{Kovacs:2016juc} (left column of Table~\ref{Tab:param}), while in the bottom ones the parameter set from \cite{Kovacs:2021ger} (right column of Table~\ref{Tab:param}) is used. The inset shows the divergence of the $Z_\pi$ wave function renormalization constant, which give rise to the divergence of the pion and eta masses.}    
        \label{fig:meson_masses}
\end{figure*}

Besides the condensates, one can also check the $N_c$ scaling of the tree-level meson masses. The explicit expressions of all the meson masses can be found in \cite{Kovacs:2016juc} and \cite{Parganlija:2012fy}. 
The $N_c$-dependence of the meson masses are shown in Fig.~\ref{fig:meson_masses}. 
In the top two figures the parameter set of the left column of Table~\ref{Tab:param} is used. On the left the scalar and pseudoscalar masses,  while on the right the vector and axial vector masses can be seen. For most of the masses---except for the pion and eta---after some transient ---$N_c \lessapprox 20$---the expected $N_c^0$ behaviour sets in.

The changes at low $N_c$ has two sources, the first one is due to the changes of the condensates for low values of $N_c$ and the second is due to the subleading terms in the masses. Both disappear rather quickly with increasing $N_c$. This can be demonstrated for example in the case of the $a_1$ mass, which is given by\footnote{In the expression below all parameters should be regarded as $N_c$ dependent (see Table~\ref{tab:Nc_dep})}. 
\be \begin{split}
m_{a_1}^2 =& m_1^2 + \frac{h_1}{2}\phi_S^2 +2\delta_N \\ &+ \frac{1}{2} \left( 2 g_1^2 +h_1+h_2-h_3\right) \phi_N^2
,
\end{split} \ee
where every term has a $N_c^0$ scaling except for the $h_1\phi_N^2/2$ term, which is $\propto N_c^{-1}$. This term will cause a drop (or rise if it is negative) in the mass as it vanishes with the increasing value of $N_c$. Such a term appears in each vector and axial-vector masses with $h_1$ and in each scalar and pseudoscalar masses with $\lambda_1$. If the initial value of $h_1$ or $\lambda_1$ is relatively large, then these terms give a major contribution to the masses and the change up to $N_c \lessapprox 20$ will be also significant.

Due to the large change of the vector and axial vector masses, a problem may arise, as it can be seen in the top left figure in Fig.~\ref{fig:meson_masses}:
the mass of the pion and the $\eta_L$ fields increase---and eventually diverges at around $N_c\approx 400$. This unwanted behavior is a consequence of the mixing between the axial vector and pseudoscalar sector, viz., the masses of the pseudoscalars contain a wavefunction renormalization factor, which can have a zero denominator for certain values of the parameters. This $Z_{\pi}$ factor for the pion and the $\eta_N$ (the non-strange part of the pseudoscalar-isoscalar sector) reads as
\be \label{Zpi}
Z_\pi = \frac{m_{a_1}}{\sqrt{m_{a_1}^2-g_1^2 \phi_N^2}},
\ee  
which is well defined only for $m_{a_1}^2 > g_1^2 \phi_N^2$. This is always true for $N_c=3$, however, if the drop---as $N_c$ increases---in the $m_{a_1}$ mass is too large, a divergence in $Z_\pi$ and thus in $m_{\pi}/m_{\eta_N}$ may appear. As mentioned above this behavior of the $m_{a_1}$ mass follows from that the value of $h_1$ at $N_c=3$ is relatively large and thus it gives a major contribution to the mass. It turns out that a relatively large $h_1$ value shows up for most of the parameter sets that provides low mass for the sigma (or $f_0$) field, which is needed to get a first order phase transition along the $\mu_B$ axis. It is interesting that, upon demanding Eq.~\eqref{Eq:NS_cond} in the parameterization, the divergence can be avoided. Such a parameter set can be seen in the right column of Table~\ref{Tab:param}. The lower figures in Fig.~\ref{fig:meson_masses} are made with this set of parameters and obviously the divergence is absent. It should be noted, however, that even without imposing Eq.~\eqref{Eq:NS_cond} we were able to find parameter sets that are free from such divergences. This shows that the condition is sufficient but not necessary.

In the future, one may consider the addition of a four-quark nonet into the PLeLSM. Namely, the light scalar state $f_{0}(500)$ as well as the other scalars below 1 GeV do not fit into the quark-antiquark picture, see the reviews \cite{Pelaez:2015qba,Klempt:2007cp} and refs. therein,  As shown in a study of the eLSM in the vacuum  \cite{Lakaschus:2018rki}, when a  light four-quark state is added, a small $h_1$ parameter is quite natural, thus no divergence would appear. In turn, the light non-conventional meson $f_0(500)$ has also shown to be potentially relevant at nonzero temperature \cite{Heinz:2008cv} and at nonzero density \cite{Gallas:2011qp,Heinz:2013hza}. 
The inclusion of four-quark objects in the PLeLSM is then a straightforward extension of the model at $N_c=3$. Yet, it is not expected to affect the large-$N_c$ results, because four-quark states disappear in this limit. 

\section{Polyakov-loops at large $N_c$}
\label{sec:polyakov_Nc}


The extension of the Polyakov-loop variables from $N_c=3$ to $N_c>3$ is a complicated tasks that requires several steps.
In this section we present a way to tackle this problem. 

The Polyakov-loop is a special Wilson-line in the temporal direction at nonzero temperature---usually periodic boundary condition is applied, hence the line becomes a loop. This provides a way to mimic the effect of confinement as it can be used to define a parameter to signal center symmetry breaking. The Polyakov-loop is defined as
\be 
\label{Eq:PolyakovL}
L(\Vec{x})=\mathcal{P}\exp\left\{i\int_0^\beta A_4 dt\right\},
\ee 
where $A_4$ is the temporal component of the gluon field in Euclidean metric and $\mathcal{P}$ the path-ordering operator. Thus, $L(x)$ is a matrix in $SU(N_c)$, which, in general, is not diagonal. To get a color singlet quantity one usually defines the color traced Polyakov-loops, or Polyakov-loop variables 
\be 
\Phi(\Vec{x}) =\frac{1}{N_c} \Tr_c L(\Vec{x}) , \; \text{and}\; \bar \Phi(\Vec{x}) =\frac{1}{N_c} \Tr_c L(\Vec{x})^\dagger,
\ee
which are gauge invariant, but not invariant under nontrivial center transformations---i.e. $\mathcal{C}=c\mathbf{1}$, $c\in\mathbb{C}$, $\lvert c\rvert\neq 1$. It is shown in \cite{McLerran:1981pb} that the thermal expectation values of the Polyakov-loop variables are related to the $\Delta F_{q/\bar q}$ change in the free energy, when an infinitely heavy quark (or antiquark) is added to the system
\be 
\langle\Phi (\Vec{x})\rangle_{\beta} = e^{-\beta \Delta F_q (\Vec{x})} \; \text{ , }\; \langle \bar \Phi (\Vec{x})\rangle_{\beta} = e^{-\beta \Delta F_{\bar q} (\Vec{x})}  \text{ .}
\ee 
Since $\Phi(\Vec{x})$ is not invariant under nontrivial center transformations, a center symmetric phase $\langle\Phi (\Vec{x})\rangle_{\beta}=0$ implies $\Delta F_q (\Vec{x})=\infty$, which means confinement. In the deconfined phase $\Delta F_q (\Vec{x}) < \infty$ and consequently $\Phi(\Vec{x})\neq 0$. Similar argument holds for $\bar \Phi(\Vec{x})$ for antiquarks. Thus $\Phi (\Vec{x})$ and $\bar \Phi (\Vec{x})$ can be used as order parameters of the phase transition between the confined and deconfined phases. It should be noted that for $\mu_q=0$ $\langle\Phi (\Vec{x})\rangle_{\beta} = \langle \bar \Phi (\Vec{x})\rangle_{\beta}$, but for $\mu_q \neq 0$ they are not equal.  

As usual, we apply the Polyakov gauge, in which $A_4$ is time independent and diagonal. As a further simplification we take an $\Vec{x}$-independent, i.e. homogeneous, gluon field. Consequently $L$ can be written as
\be \label{Eq:L_F}
    L=e^{i\beta A_4} = \diag \left( e^{iq_1},\ldots ,e^{iq_{N_c}} \right),
\ee
where $q_j \in \mathbb{R}$ are some phases and $\sum_j q_j=0$.

Among the diagonal $SU(N_c)$ matrices, there are $N_c-1$ independent elements, thus the two Polyakov-loop variables---that were defined above---are not sufficient alone to completely describe the symmetry breaking for $N_c > 3$. As it is discussed in \cite{Dumitru:2012fw,Nishimura:2017kjl}, one has to define $N_c-1$ independent quantities, for e.g., the color traced Polyakov-loops that wind $n$ times around in temporal direction,
\be 
\label{Eq:phi_N_def}
\Phi_n = \frac{1}{N_c} \Tr_c L^n, \quad \bar \Phi_n = \frac{1}{N_c} \Tr_c {L^\dagger}^n,
\ee \\
where $n\in \left(1,\dots ,\lfloor\frac{N_c}{2}\rfloor\right)$.
These objects form a complete set of order parameters. It should be noted, however, that for $N_c$ even there are $N_c/2$ variables $\Phi_n$ and $N_c/2$  variables $\bar \Phi_n$, thus altogether there are $N_c$ Polyakov-loop objects, but only $N_c-1$ of them are independent. Thus, there are a $\Phi_k$ and a related $\bar \Phi_k$ which only appear in certain combinations and cannot be determined separately. On the other hand if $N_c$ is odd, then all the $\Phi_n$ and $\bar \Phi_n$ are independent.     

The Polyakov loop variables were already introduced in the PLeLSM for $N_c=3$ in \cite{Kovacs:2016juc}. Our goal is to calculate the grand potential Eq.~\eqref{Eq:grand_pot}, in which the Polyakov-loop variables appear in the second (the fermionic part) and third (the Polyakov-loop part) terms. 
As explained in detail in Sec. III of \cite{Kovacs:2016juc}, the $\Omega_{\bar qq}^{(0)} (T,\mu_q)$  fermion part of the grand potential can be calculated from the partition function,
\be
\mathcal{Z}_{\bar q q}^{(0)} = e^{-\beta V \Omega_{\bar qq}^{(0)}},
\ee
where 
\begin{equation} 
  \Omega_{\bar q q}^{(0)}(T,\mu_q) = \Omega_{\bar q q}^{(0)\textnormal{v}} + \Omega_{\bar q q}^{(0)\textnormal{T}}(T,\mu_q),
\end{equation}
consists of a vacuum and a thermal part. In this approximation, basically, the quarks propagate on a constant gluon background, which amounts to adding a color dependent contribution the chemical potentials of the quarks in the thermal part. The calculation can be easily generalized to $N_c>3$ (see also \cite{Hansen:2006ee}) leading to 
\begin{align}
  &\Omega_{\bar q q}^{(0)\textnormal{v}} = -2 N_c \sum_{f=u,d,s} \int\frac{d^3 p}{(2\pi)^3} E_f(p), \label{Eq:fermion_vac}\\
  &\Omega_{\bar q q}^{(0)\textnormal{T}}(T,\mu_q)  = -2 T \Tr_c \sum_{f=u,d,s} \int\frac{d^3 p}{(2\pi)^3} \nonumber \\
  &\times \big[\ln\big(1+ L^\dagger e^{-\beta (E_f(p)- \mu_q)}\big)
  +\ln\big(1+ L e^{-\beta (E_f(p)+ \mu_q)}\big)\big] \nonumber\\
  &\equiv -2 T \sum_f\int
\frac{d^3 p}{(2\pi)^3}\big[\ln g_f^+(p) + \ln g_f^-(p)\big] \text{ ,} \label{Eq:fermion_thermal}
\end{align}
where we have introduced $g^{\pm}_f$,
\begin{align}
\ln g^+_f(p) &\equiv \Tr_c \ln \left[ \mathbb{1}+L^\dagger e^{-\beta(E_f(p)-\mu_q)}\right] \nonumber\\
             &= \ln \Det_c \left[ \mathbb{1} + L^\dagger  e^{-\beta (E_f(p)-\mu_q)} \right]  \label{Eq:g_pm} \text{ ,}\\
\ln g^-_f(p) &\equiv \Tr_c \ln \left[ \mathbb{1}+L e^{-\beta(E_f(p) + \mu_q )}\right] \nonumber \\
             &= \ln \Det_c \left[ \mathbb{1} + L  e^{-\beta(E_f(p) + \mu_q)} \right] \text{ .} \nonumber
\end{align}
Introducing $E_f^{\pm} \equiv E_f(p) \mp \mu_q $, the quantity $g^+_f$ can be written for a general $N_c$ as
\begin{align} 
g^+_f=&\Det_c \left[ \mathbb{1} + L^\dagger  e^{-\beta E_f^+} \right] \nonumber\\
     =& \left( 1+ e^{-iq_1} e^{-\beta E_f^+}\right)\cdots \left( 1+ e^{-iq_{N_c}}e^{-\beta E_f^+}\right) \nonumber\\
     =& 1+e^{-i\sum_j q_j}e^{-N_c\beta E_f^+} + \sum_j e^{-i q_j}e^{-\beta E_f^+} \nonumber\\
     +&\sum_j e^{-i\sum_{k\neq j} q_k}e^{-(N_c-1)\beta E_f^+} \nonumber \\
     +& [\text{terms with $2$ to $N_c-2$ phases}], \label{Eq:gf+Nc1}  
\end{align} 
where we have separated terms that contain $0$, $N_c$, $1$, and $N_c-1$ number of $q_j$ phases. The remaining terms contain minimum $2$, maximum $N_c-2$ number of phases. One can use that $\sum_j q_j=0$, $\sum_{k\neq j} q_k =-q_j$ and the Eq.~\eqref{Eq:phi_N_def} definition of the Polyakov-loop variables to get
\begin{align} 
g^+_f =& 1+e^{-N_c\beta E_f^+} + N_c\left[\bar \Phi_1 e^{-\beta E_f^+} + \Phi_1 e^{-(N_c-1)\beta E_f^+}\right] \nonumber\\
      +& [\text{terms with 2 to Nc-2 phases}] \text{ ,}
\end{align} 
where the first line agrees with the result in \cite{Hansen:2006ee} and \cite{Kovacs:2016juc} for $N_c=3$, in which case the last line vanishes. 

The expression for $g^-_f$ is given by interchanging $\Phi_1$ and $\bar \Phi_1$ and changing $E_f^+$ to $E_f^-$ in $g^+_f$. For $N_c>3$ not only $\Phi_1$ and $\bar \Phi_1$ but also other $\Phi_k$s will appear. One can calculate for instance terms that contain $2$ and $N_c-2$ different phases. After some calculations they read 
\begin{align}
&\frac{1}{2}\left( N_c^2 \bar \Phi_1^2 - \Tr_c {L^\dagger}^2 \right) e^{-2\beta E_f^+},\; \text{for $2$ phases,} \\
&\frac{1}{2}\left( N_c^2 \Phi_1^2 - \Tr_c {L}^2 \right) e^{-(N_c-2)\beta E_f^+},\; \text{for $N_c-2$ phases,}
\end{align}
where $\Tr_c {L}^2$ and $\Tr_c {L^\dagger}^2$ cannot be expressed with $\Phi_1$ and $\bar \Phi_1$. According to Eq.~\eqref{Eq:phi_N_def} , $\Phi_2$ and $\bar \Phi_2$ appear:  
\begin{align}
  g^+_f =& 1+e^{-N_c\beta E_f^+} \nonumber\\
        +& N_c\left[\bar{\Phi} e^{-\beta E_f^+}+\Phi e^{-(N_c-1)\beta E_f^+}\right]\nonumber \\ 
        +& \frac{1}{2}\left( N_c^2 \bar \Phi^2 - N_c \bar \Phi_2 \right) e^{-2\beta E_f^+} \\ \label{Eq:general_det_dagger_upto2}
        +& \frac{1}{2}\left( N_c^2 \Phi^2 - N_c \Phi_2 \right) e^{-(N_c-2)\beta E_f^+}\nonumber\\
        +& [\text{terms with 3 to Nc-3 phases}] \text{ .}\nonumber
\end{align}
With increasing $N_c$ more and more new unknown $\Phi_k$ and $\bar \Phi_k$ also emerge\footnote{A new $\Phi_k$ appears whenever $N_c=2k$.}. 
Accordingly, as $N_c$ increases more and more Polyakov-loop variables or order parameters are needed, and consequently the number of field equations to be solved is also increasing. At a given $N_c$ there are $N_c+1$ field equations, thus the task is not feasible already for $N_c\gtrsim 10$. Consequently, we need a reasonable approximation, which can drastically reduce the number of independent Polyakov-loop variables. 

\subsection{Uniform eigenvalue Ansatz}

To reduce the degrees of freedom in the Polyakov sector of the model we will use the so called uniform eigenvalue Ansatz (UEA), which was defined in \cite{Dumitru:2012fw} using group theoretical considerations and was already used recently in \cite{Lo:2021qkw}, where the deconfinement phase transition was investigated for $N_c=2,3,4$ in an $SU(N_c)$ effective model approach. Within this Ansatz the $q_j$ phases in the $L$ operator can be written as
\be 
q_j(s)=-\pi\frac{N_c-2j+1}{N_c}s, \quad 0\leq s\leq 1, \quad j \in 1\ldots N_c
\label{Eq:qj_w_s}
\ee 
where the confining ($L=\mathbb{1}$) and the perturbative ($L=0$) vacua correspond to the points $s=0$ and $s=1$, respectively\footnote{We stress that there is a difference between the notation of the current work and the one of \cite{Dumitru:2012fw}, where a $2\pi$ is factored out from $\vec q$.}. Here, the variable $s$ plays a role of some external parameter (like $\beta$). Notice that, in this approximation, the eigenvalues of the Polyakov loop operator are points on the unit circle with equally distributed angles, while in the $N_c\to \infty$ limit gives a uniform eigenvalue density for $L$ in the range of $(-\pi s, \pi s)$. 

To implement the UEA in our model one may express---similarly to \cite{Lo:2021qkw}---the general $q_j$ angles of Eq.~\eqref{Eq:L_F} in terms of the $N_c-1$ group angles of the Cartan subgroup of $SU(N_c)$
\be  \label{Eq:def_group_angles}
\vec q \equiv (q_1,\ldots, q_{N_c}) = \sum_{j=1}^{N_c-1} \gamma_j \vec v_j \text{ ,}
\ee
with $\lbrace \vec v_j\rbrace_{j=1}^{N_c-1}$ being a set of basis vectors. Their $N_c$ number of elements sum up to zero to fulfill the condition $\sum_j q_j=0$ coming from the special unitarity. These basis vectors can be written in such a way that the elements of $v_1$ are equidistant,
\begin{align}
    N_c &= 3 \quad \vec v_1 = \left( -1,0,1 \right),\nonumber \\
    N_c &= 4 \quad \vec v_1 = \left( -1,-1/3,1/3,1 \right),\nonumber\\
    N_c &= 5 \quad \vec v_1 = \left( -1,-1/2,0,1/2,1 \right),\nonumber \\
    &\vdots \nonumber\\
    N_c, &\quad \vec v_1 = \left( -1,-\left(1-\frac{2}{N_c-1}\right),\ldots,\right.\nonumber\\
    &\left. \hfill -\left(1-(j-1)\frac{2}{N_c-1}\right),\ldots ,1 \right), \; j=1,\ldots,N_c \text{ .}\label{Eq:v1_Nc}
\end{align} 
It is clear that, keeping as nonzero only the coefficient of this vector, i.e. $\gamma_1\neq 0$, $\gamma_i =0, i\ne 1$, corresponds to the uniform eigenvalue Ansatz. This means that a single direction is fixed in the Cartan subalgebra (of $SU(N_c)$) and the Polyakov loop is calculated in this subspace. Accordingly, the Polyakov loop can be written with the help of a single $\gamma (\equiv \gamma_1 )$ parameter as
\begin{align}
L&= \diag \left(e^{-i\gamma},e^{-i\left(1-\frac{2}{N_c-1}\right)\gamma},e^{-i\left(1-2\frac{2}{N_c-1}\right)\gamma}, \ldots,\right.\nonumber \\ 
& \left.(e^{0}),\ldots, e^{i\left(1-2\frac{2}{N_c-1}\right)\gamma},e^{i\left(1-\frac{2}{N_c-1}\right)\gamma},e^{i\gamma}\right),
\end{align}
where $e^0$ is part of the sequence only if $N_c$ is odd. 
For example, for $N_c=6$ and $7$ one has 
\begin{align} 
    L  &=\diag \left(e^{-i\gamma},e^{-i3\gamma/5},e^{-i\gamma/5},e^{i\gamma/5},e^{i3\gamma/5},e^{i\gamma}\right),\nonumber\\
    &\text{and}\label{Eq:L_67} \\
    L  &=\diag \left(e^{-i\gamma},e^{-i2\gamma/3},e^{-i\gamma/3},e^{0},e^{i\gamma/3},e^{i2\gamma/3},e^{i\gamma}\right),\nonumber
\end{align}
respectively. Plugging Eq.~\eqref{Eq:v1_Nc} into Eq.~\eqref{Eq:def_group_angles} and comparing to Eq.~\eqref{Eq:qj_w_s}, the connection between $\gamma$ and $s$ is given by $\gamma =\pi\frac{N_c-1}{N_c}s$. 

It can be seen that the color trace of each power of $L$ is real, thus, the Polyakov loop variables 
are also such, i.e. $\Phi_n = \bar \Phi_n \in \mathbb{R}$ for each $n$, and can be written explicitly as
\be 
\Phi_n =\frac{1}{N_c} \left(2\sum_{j=1}^{\lfloor\frac{N_c}{2}\rfloor} \cos\left[ \left(1-2\frac{j-1}{N_c-1}\right)n\gamma\right] + \alpha \right)
\ee 
with $\alpha = 1$ if $N_c$ is odd, and $\alpha = 0$ if $N_c$ is even. It is important to stress that in the UEA approximation everything that is necessary for the calculation of the grand potential can be expressed with the help of $\gamma$. Actually, it turns out that, instead of $\Phi$, it is much easier to work directly with $\gamma$ and calculate any $\Phi_n$ afterwards.

At vanishing chemical potential this approximation is exact for $N_c\leq 3$ in effective models with $Z(3)$ and $\Phi\leftrightarrow\bar \Phi$ symmetric Polyakov-loop potential, for which the Polyakov-loop variables are real for any temperature. Moreover, within the matrix model \cite{Dumitru:2012fw} it gives less than one percent deviation from the exact solution for $N_c=7$. 
At finite $\mu_q$  the Polyakov loop variables are already complex for $N_c\geq 3$ ($\Phi_3=\Phi\neq \bar \Phi= \bar \Phi_3$). At $N_c=3$, one can test validity of the Ansatz above by comparing the thermodynamics of a model where the Ansatz is used and with another model with two (i.e. $N_c-1$) independent Polyakov-loop variables like in \cite{Kovacs:2016juc} (see later). Moreover, in order to use the uniform eigenvalue Ansatz within the PLeLSM, we also need a Polyakov loop potential, which is compatible with this approximation and applicable at $N_c>3$.

\subsection{Polyakov-loop potential at large $N_c$} 
\label{Ssec:Polyakov_Nc}

The Polyakov-loop potential, taken from \cite{Lo:2021qkw}, reads\footnote{Note, in \cite{Lo:2021qkw} three possible forms are given; we use here their model A.}  
\be \label{Eq:UPol_uea}
U_\Pol =U_\conf+U_\glue,
\ee 
where the two terms refer to a confining and a deconfining part. The first term of the potential is explicitly given by
\begin{equation}
\label{Eq:Uconf_uea}
  U_\conf = -\frac{b}{2}T\ln H
\end{equation}  
with $H$ being the invariant Haar-measure of the $SU(N_c)$ group and $b$ is a parameter. The potential in \eqref{Eq:Uconf_uea} is confining in the sense that it has a minimum in the center symmetric vacuum at $\Phi=0$. 
The Haar-measure is expressed via the $q_j$ phases as:
\begin{equation}
    H=\prod_{j>k}\left| e^{iq_j} - e^{iq_k} \right|^2 =\prod_{j>k} 4\sin^2 \left(\frac{q_j-q_k}{2} \right).
\end{equation} 
 
The deconfinement is induced by the second term in \eqref{Eq:UPol_uea}, which is given by
\begin{align}
  U_\glue &= n_\glue T \int \frac{d^3p}{(2\pi)^3} \Tr \ln \left( \mathbb{1}_A - L_A e^{-\beta E_A(p)}\right) \nonumber \\ 
  &=n_\glue T \int \frac{d^3p}{(2\pi)^3} \ln \Det \left( \mathbb{1}_A - L_A e^{-\beta E_A(p)}\right) \nonumber\\
  &\equiv n_\glue T \int \frac{d^3p}{(2\pi)^3} \ln g_A \text{ ,} \label{Eq:g_A}
\end{align}  
where $g_A = \Det \left( \mathbb{1}_A - L_A e^{-\beta E_A(p)}\right)$, $E_A(p)=\sqrt{p^2+m_A^2}$, $\mathbb{1}_{A}$ is the $N_c^2-1$ dimensional identity, and $L_A$ is the Polyakov-loop operator in the adjoint representation:
\be \label{Eq:L_A}
L_A=\diag (e^{iQ_1},\ldots,e^{iQ_{N_c^2-1}})
\ee 
with the $Q_j$ adjoint angles, 
\be \label{Eq:adjoint_angle}
\vec Q = (\underbrace{0,\ldots,0}_{N_c-1},\underbrace{q_1-q_2,\ldots,(q_j-q_k)|_{j\ne k},\ldots q_{N_c-1}-q_{N_c}}_{N_c(N_c-1)}).
\ee 
The adjoint angles are constructed from the root system \cite{Georgi:1999wka, Hall:Lie, Dumitru:2012fw, Lo:2021qkw} and classified into a Cartan part with $N_c-1$ zeros and a non-Cartan part with $N_c(N_c-1)$ phase differences.
Moreover, there are three unknown parameters in the potential, namely $b$, $n_{\glue}$ (the multiplicity of the gluon field), and $m_A$ (the effective gluon mass). Their values are taken from \cite{Lo:2021qkw}, namely,  $b=(0.1745~\GeV)^3$, $n_\glue=2$, and $m_A = 0.756~\GeV$.

It is useful to express $\ln H$ in terms of the adjoint operator $L_A$ as
\begin{align}
  \label{Eq:H_Tr'}
  \ln H &=  \Tr^{\prime} \ln \left( \mathbb{1}_A - L_A \right) = \ln \Det^{\prime} \left( \mathbb{1}_{A} - L_A \right),\\
  & \equiv \ln g^{\prime}_A \label{Eq:g_a_pr}
\end{align}
where $g^{\prime}_A$ is introduced and the prime on the $\Tr$ and $\Det$ denotes a partial trace/determinant over the non-Cartan roots only.

Besides the UEA described above, we have also employed another approximation, according to which we assumed that $\Phi_n=\Phi^n$. In this approximation the Polyakov loop variables are complex for $\mu_q>0$, and this approximation gives very similar results as the UEA. Some details of this approach can be found in Appendix~\ref{App:phi_n_phi_n}. The qualitative picture emerging in the large-$N_c$ limit remains unchanged.

\subsection{Field equations in the UEA approximation}
\label{Ssec:FE_uea}

The grand potential of Eq.~\eqref{Eq:grand_pot} consist of three terms, from which the first mesonic part ($U(\langle M\rangle)$) is given explicitly in Eq.~(20) of \cite{Kovacs:2016juc}. The second fermionic term has a vacuum and a matter part. The vacuum part (Eq.~\eqref{Eq:fermion_vac}) needs to be renormalized (see  the explicit expression in in Eq.~(31) of \cite{Kovacs:2016juc}). For the matter or thermal part (Eq.~\eqref{Eq:fermion_thermal}), $\ln g_f^{\pm}$ is needed in the UEA approximation. For this we start from Eqs.~\eqref{Eq:g_pm}-\eqref{Eq:gf+Nc1}  and use the fact that phase factors always appear in pairs, i.e same absolute value, but opposite sign (see e.g. Eq.~\eqref{Eq:L_67}), thus a general term can be written as    
\begin{align}
    &\left(1 + e^{-i\theta}e^{-\beta E_f^{\pm}} \right)\left(1 +  e^{i\theta}e^{-\beta E_f^{\pm}} \label{Eq:gf_pairs} \right)\nonumber \\
&= 1 + 2\cos(\theta) e^{-\beta E_f^{\pm}}+ e^{-2\beta E_f^{\pm}}, 
\end{align}
and there are $\lfloor N_c/2 \rfloor$ number of such pairs. In case of $N_c$ being odd, there is an extra $(1 + e^{-\beta E_f^{\pm}})$ factor compared to $N_c$ being even. Moreover, it is clear that in the UEA approximation $g_f^{\pm}$ is real and its explicit form reads as
\begin{align}   \nonumber
g^\pm_f =& \prod_{j=1}^{\lfloor N_c/2 \rfloor} \left(1+2\cos (\theta_j) e^{-\beta E^\pm_f} +e^{-2\beta E^\pm_f} \right) \\  \nonumber
& \qquad\quad \times \begin{cases}
 \left(1+e^{-\beta E^\pm_f}\right) & \text{for $N_c$ odd} \\  \nonumber
 1 & \text{for $N_c$ even}
\end{cases},\label{Eq:gf_pm_expl} \\
\end{align}
where $\theta_j$ are the $\lfloor N_c/2 \rfloor$ different positive angles, i.e. the positive $q_j$'s 
\be 
\theta_j = \left(1- (j-1)\frac{2}{N_c-1} \right) \gamma \quad j=1\ldots \lfloor N_c/2 \rfloor.
\ee 
The $g_A$ and $g^{\prime}_A$ determinants, introduced in Eq.~\eqref{Eq:g_A} and Eq. \eqref{Eq:g_a_pr}, are also real and can be formulated in a similar fashion as $g_f^{\pm}$:
\begin{align}
g_A=& \prod_{j=1}^{N_c(N_c-1)/2} \left(1-2\cos (\mathcal{Q}_j) e^{-\beta E_A(p)} +e^{-2 \beta E_A(p)} \right)  \nonumber \\ 
&\times \left(1-e^{-\beta E_A(p)}\right)^{N_c-1},\label{Eq:g_AApr}\\
g^{\prime}_A=& \prod_{j=1}^{N_c(N_c-1)/2} 2 \left( 1- \cos(\mathcal{Q}_j) \right),
\end{align}
where $\mathcal{Q}_j$ are the $N_c(N_c-1)/2$ angles calculated as the differences of two different $q_j$'s, i.e.  $\mathcal{Q}_j \in \left\{ q_{k} - q_{l} | k,l=1\ldots N_c; k<l \right\}$.

Consequently, the Polyakov loop dependent part of the grand potential of Eq.~\eqref{Eq:grand_pot} (second and third terms), reads as:
\begin{align}    \label{Eq:Grand_uea_Pol}
\left. \Omega\right|_\Pol =& U_\Pol + \Omega_{\bar qq}^\text{matter} =\\ \nonumber
&-\frac{b}{2} T \ln g_A' + n_\glue T \int \frac{d^3p}{(2\pi)^3} \ln g_A
\\ \nonumber
&-2 T \sum_{f=u,d,s} \int \frac{d^3p}{(2\pi)^3} \left[\ln g^+_f +\ln g^-_f \right]. \nonumber 
\end{align} 

As it was mentioned above, by using the uniform eigenvalue Ansatz one can express the Polyakov-loop variable dependence as a function of solely the phase variable $\gamma$, that gives rise to an additional field equation beside those for the meson condensates,
\be 
0=\frac{\partial \Omega}{\partial\gamma}= \frac{\partial U_\Pol}{\partial\gamma} +\frac{\partial \Omega_{\bar qq}^\text{matt}}{\partial\gamma}.
\ee 
The field equations for the meson condensates are modified only in the matter part of the fermion integral, which can be written as
\be \label{Eq:dOmperdphiX}
\frac{\partial\Omega_{\bar q q}^\matt}{\partial \phi_{N/S}} = -4N_c \sum_{f=u,d,s} m_f \frac{\partial m_f}{
\partial \phi_{N/S}} \mathcal{T}_f^\matt \text{ .}
\ee  
where the matter part of the tadpole integral\footnote{The factor of 4 in Eq.~\eqref{Eq:dOmperdphiX} is written to match the definition of the tadpole integral of Eq.~(28) in \cite{Kovacs:2021kas} for $N_c=3$ with a modified Fermi-Dirac distribution.} is
\be  \label{Eq:def_Tmatt_uea}
\mathcal{T}_f^\matt = \frac{T}{N_c} \int \frac{d^3 p}{(2\pi)^3} \frac{1}{2 m_f} \left( \frac{1}{g_f^+} \frac{\partial g_f^+}{\partial m_f} + \frac{1}{g_f^-} \frac{\partial g_f^-}{\partial m_f}\right).
\ee 
Finally, the three field equations are
\begin{widetext}
\begin{subequations} \label{Eq:Field_equations_UEA}
\begin{align}
 0 &= \frac{d U_\text{Pol}}{d\gamma} -2T \sum_{f=u,d,s} \int \frac{d^3p}{(2\pi)^3}  \left(h^+_f(p) +h^-_f(p) \right), \label{Eq:FE_gamma} \\
0 &=  m_0^2 \phi_N + \left(\lambda_1 + \frac{\lambda_2}{2} \right)\phi_N^3 + \lambda_1\phi_N\phi_S^2 -\frac{c}{\sqrt{2}}\phi_N \phi_S -h_{0N} + \frac{g_F }{2} \sum_{l=u,d} \langle \bar{q_l} q_l\rangle_T, \label{Eq:FE_Phi_N} \\
0 &=  m_0^2 \phi_S + \left(\lambda_1 + \lambda_2 \right)\phi_S^3 + \lambda_1\phi_N^2\phi_S -\frac{\sqrt{2}c}{4}\phi_N^2 -h_{0S} + \frac{ g_F}{\sqrt{2}} \langle \bar{q_s} q_s\rangle_T, \label{Eq:FE_Phi_S} 
\end{align}
\end{subequations}
\end{widetext}
where, using the explicit form of $g_f^\pm$ given in Eq.~\eqref{Eq:gf_pm_expl}),
\begin{align}
h^\pm_f =& \frac{1}{g_f^\pm } \frac{\partial g_f^\pm }{\partial \gamma} \nonumber \\
=& \sum_{j=1}^{\lfloor N_c/2 \rfloor} \frac{ -2\sin (\theta_j) e^{-\beta E^\pm_f} \left(1- (j-1)\frac{2}{N_c-1}\right)}{1+2\cos (\theta_j) e^{-\beta E^\pm_f} +e^{-2\beta E^\pm_f} },
 \end{align}
 which is the modified Fermi-Dirac distribution function for arbitrary $N_c$ in case of the uea approximation. Moreover, the $f$ flavored constituent quark tadpole is given by
 \be 
 \langle\bar{q}_f q_f \rangle_T = -4 N_c m_f \left[  \frac{m_f^2}{16\pi^2}\left(\frac{1}{2}+\ln \frac{m_f^2}{M_0^2}\right)+ \mathcal{T}_f^\matt\right]  
 \ee 
with the matter part of the tadpole defined in \eqref{Eq:def_Tmatt_uea}.

\section{Results}
\label{sec:results}

\subsection{Zero temperature}
\label{Ssec:res_zero_T}

As already mentioned, at zero temperature the Polyakov-loop variables are exactly zero, therefore 
at $T=0$ one has to solve only Eq.~\eqref{Eq:FE_Phi_N} and \eqref{Eq:FE_Phi_S} from the system of equations of Eq.~\eqref{Eq:Field_equations_UEA}.   

\begin{figure}[htbp]
    \centering
    \includegraphics[width=0.48\textwidth]{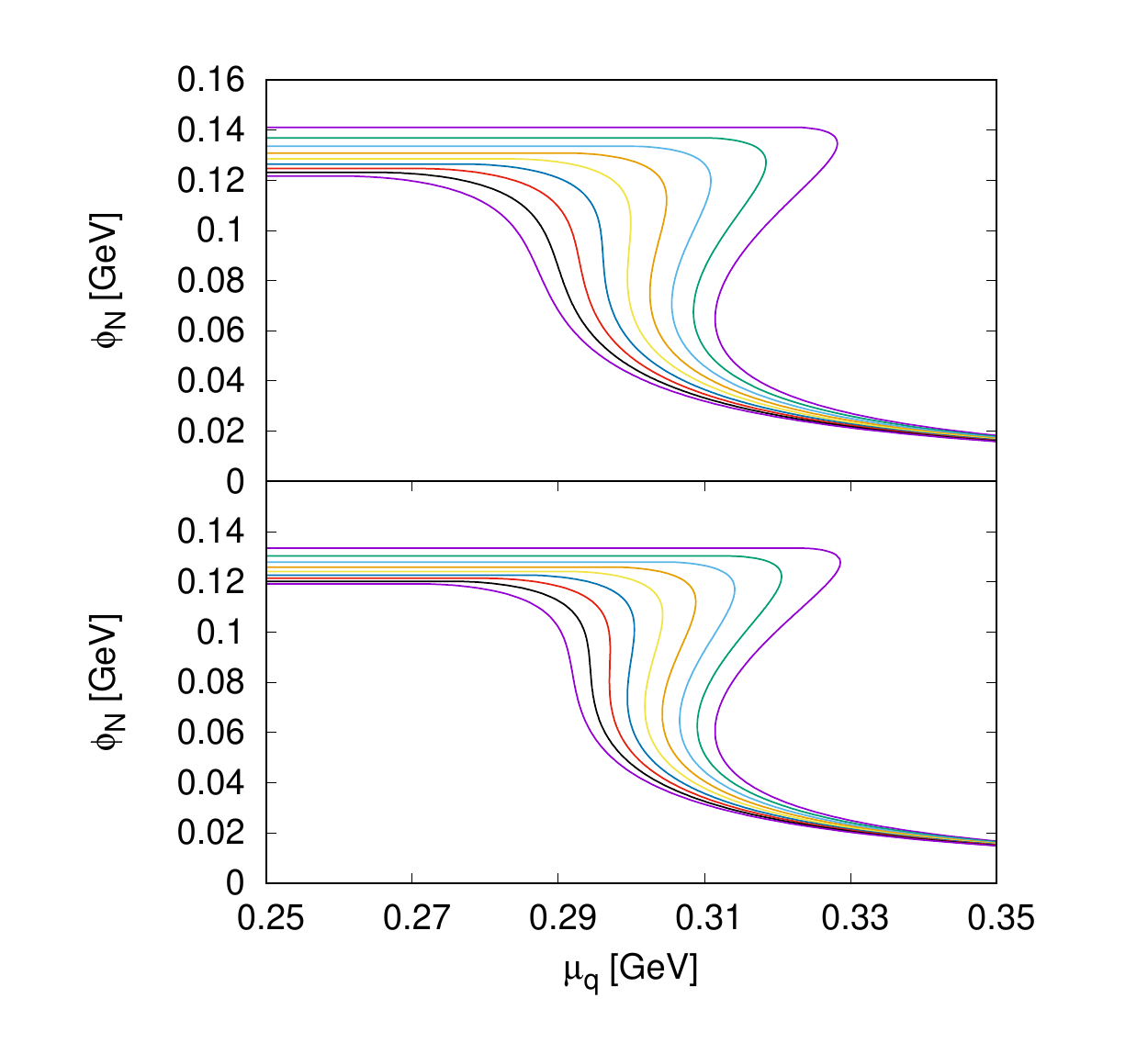}
    \caption{The $\mu_q$ quark chemical potential dependence of the $\phi_N$ condensate at different $N_c$ values. $N_c=3.00$ corresponds to the rightmost, while $N_c=3.45$ corresponds to the leftmost curve. The top figure is obtained with set A, while the bottom figure with set B of Table~\ref{Tab:param}.}
    \label{fig:phiN_muB_zeroT}
\end{figure}

We find that, upon using a parameter set that produces a first order phase transition along the $\mu_q$ axis---both parameter sets in Table~\ref{Tab:param} are like that---, the first order transition turns into a crossover already for $N_c=4$. In Fig.~\ref{fig:phiN_muB_zeroT} it is shown how the first order phase transition transforms into a crossover for both parameter sets---here for illustrative purposes $N_c$ is treated as a continuous variable.
If the transition is of crossover type then there is no CEP. Basically with increasing $N_c$ the CEP goes toward the $\mu_q$ axis and disappears, this happens already around $N_c \approx 3.3$.

\subsection{Finite temperature and zero quark chemical potential}

\begin{figure}[htbp]
    \centering
    \includegraphics[width=0.45\textwidth]{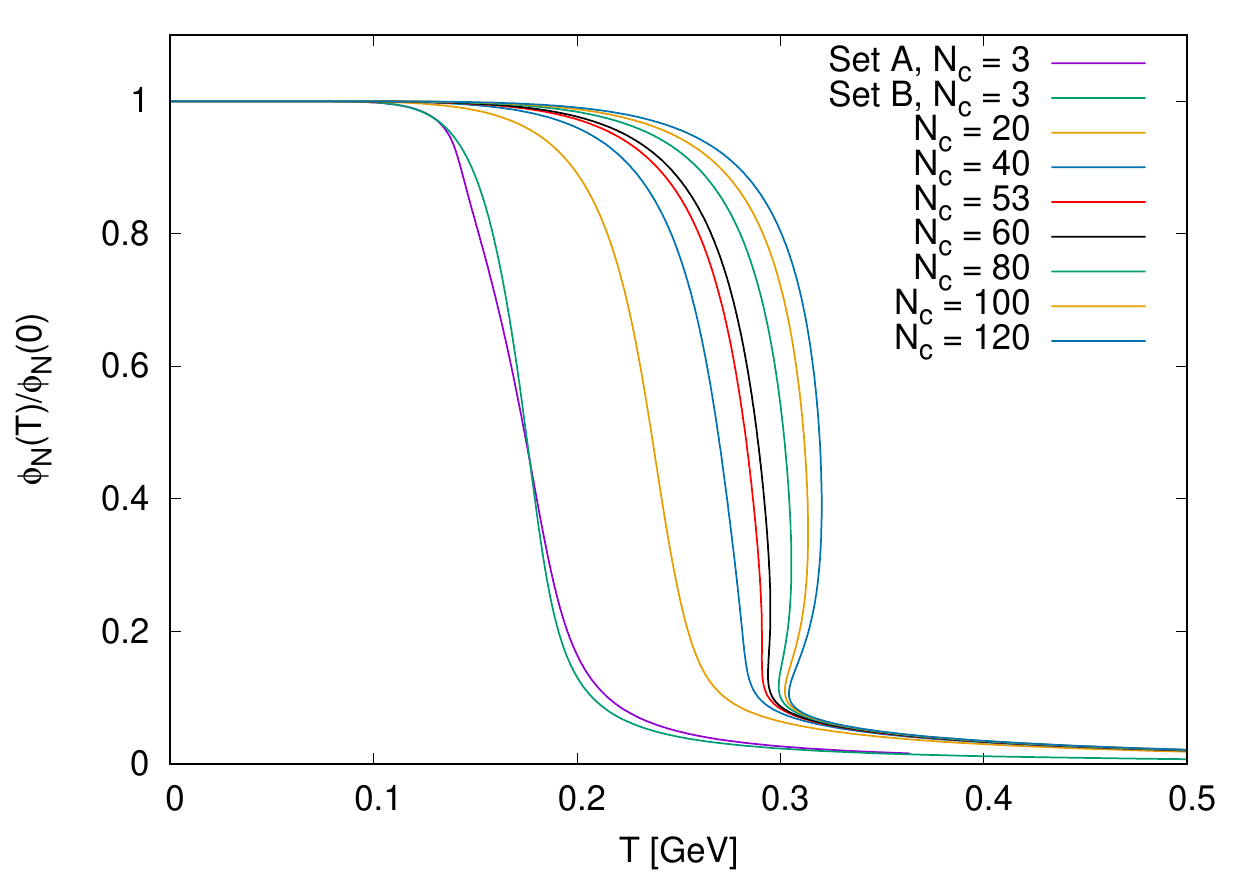}
    \caption{The temperature dependence of the normalized chiral condensate $\phi_N$.}
    \label{Fig:ops_Fuea}
\end{figure}

 At $T\ne 0$, $\mu_q = 0$ all three equations of Eq.~\eqref{Eq:Field_equations_UEA} are nontrivial. Solving them for different $T$ values starting from $T=0$ and using the solution of the three unknowns $\gamma$, $\phi_N$ and $\phi_S$ at $T-\Delta T$ as initial values at $T$ the $T$ dependence of the condensates, the masses, and all the thermodynamic variables can be determined at different $N_c$ values.  From now on, if not said otherwise, all results are made using parameter set B of Tab.~\ref{Tab:param}.  
First, we investigate the temperature dependence of the $\phi_N$ condensate for different $N_c$ values. In Fig.~\ref{Fig:ops_Fuea} the normalized $\phi_N$ is depicted for $N_c = 3, 20, \ldots 120$.   
Actually, for $N_c=3$ there are two curves, one made using Set A of  Tab.~\ref{Tab:param} and with a Polyakov-loop potential used in \cite{Kovacs:2016juc} and the other one with Set B of  Tab.~\ref{Tab:param} and with the Polyakov-loop potential described in Section \ref{Ssec:Polyakov_Nc}. As it can be seen the difference between the two curves at $N_c=3$ is very small, the values of the pseudocritical temperatures are, 
$T_c=178.6$~MeV and $T_c =176.9$~MeV, for set A and set B, respectively. 

\begin{figure}[htbp]
    \centering
    \includegraphics[width=0.45\textwidth]{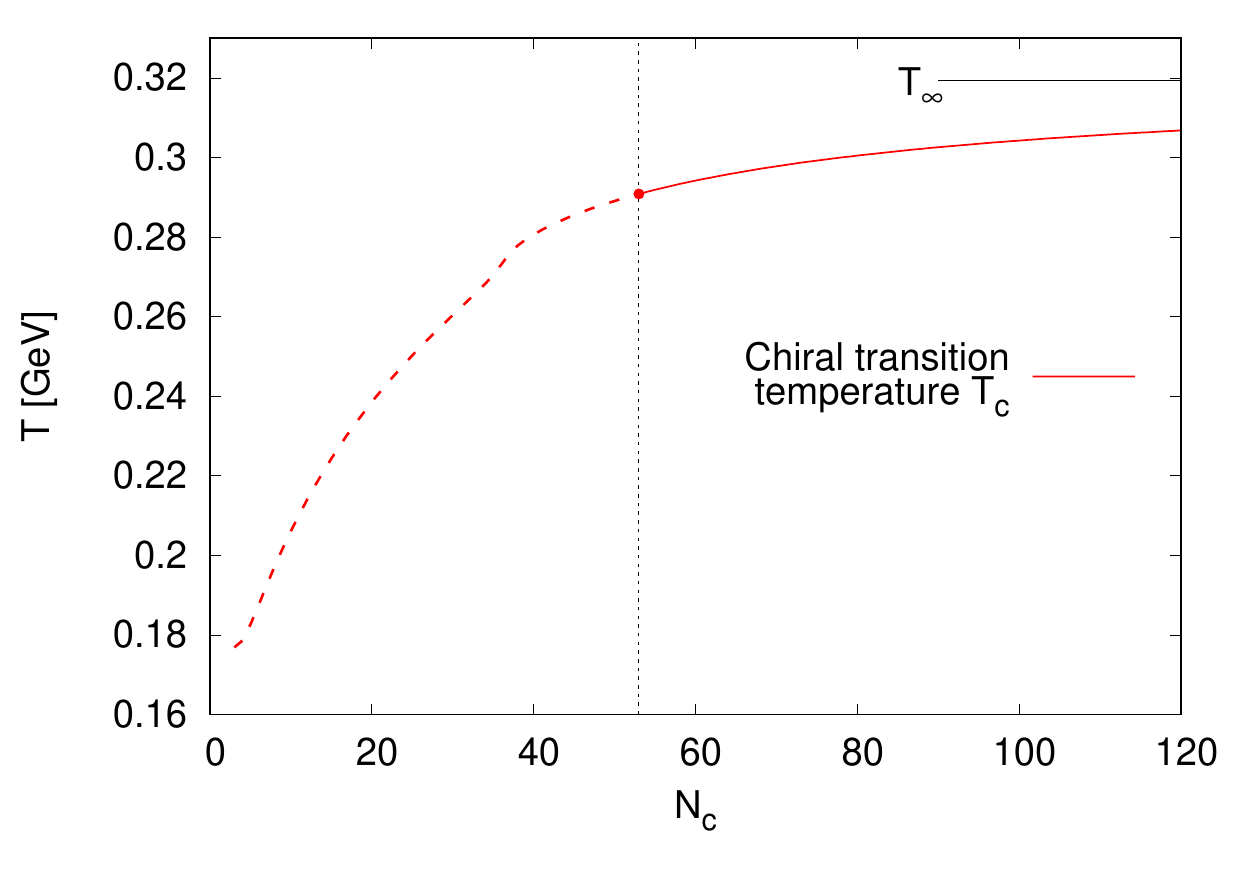}
    \caption{The $N_c$ dependence of the $T_c$ (pseudo)critical temperature.}
    \label{fig:Tc_Fuea}
\end{figure}

It can also be seen that the crossover becomes more and more abrupt with increasing $N_c$ and it eventually turns into a first order phase transition already at $N_c=53$. The curves get close to each other for increasing $N_c$, which signals a saturation in the (pseudo)critical temperature. This can be seen directly in Fig.~\ref{fig:Tc_Fuea}, where the $T_c$ (pseudo)critical temperatures are shown for different $N_c$ values.
These calculations were done only at integer values of $N_c$ (the discrete points are connected to guide the eye). The transitions are of crossover type on the left hand side and of first order type on the right hand side of the dashed line at $N_c=53$. It is worth to note that the shoulder around $N_c\approx35$ is due to the change of the dominant term in the grand potential from $\Omega_{\bar qq}^{\text{matt}}$ (scaling as $N_c$) to $U_{\Pol}$ (scaling as $N_c^2)$. We note that this change of the dominant term in the effective potential naturally leads to the appearance of a first order transition at $\mu_q=0$ for large $N_c$, independently of the actual choice of the form of the Polyakov potential. The Polyakov-loop potential alone supports a first order transition, that is melted to a crossover by the continuous behavior of the mesonic and fermionic contributions to the grand potential at $N_c=3$, as it can be seen in \cite{Fukushima:2003fw}. 

For sufficiently large $N_c$, the Polyakov sector becomes dominant and forces 
the chiral condensate $\phi_{N/S}$ to develop a first order transition. To find the $N_c\to \infty$ limit of the critical temperature one may fit the  function $T_c(N_c)=\alpha/(N_c+\beta)+T_\infty$ in the range $N_c=53\ldots 100$, where the first order transition starts and already the $U_\Pol$ dominates the grand potential. This fit gives $\alpha=-1.4865$, $\beta=-0.5554$ and $T_\infty=0.3192$~MeV, which is also shown in Fig.~\ref{fig:Tc_Fuea} with the horizontal line.

\begin{figure}[htbp]
    \centering
    \includegraphics[width=0.45\textwidth]{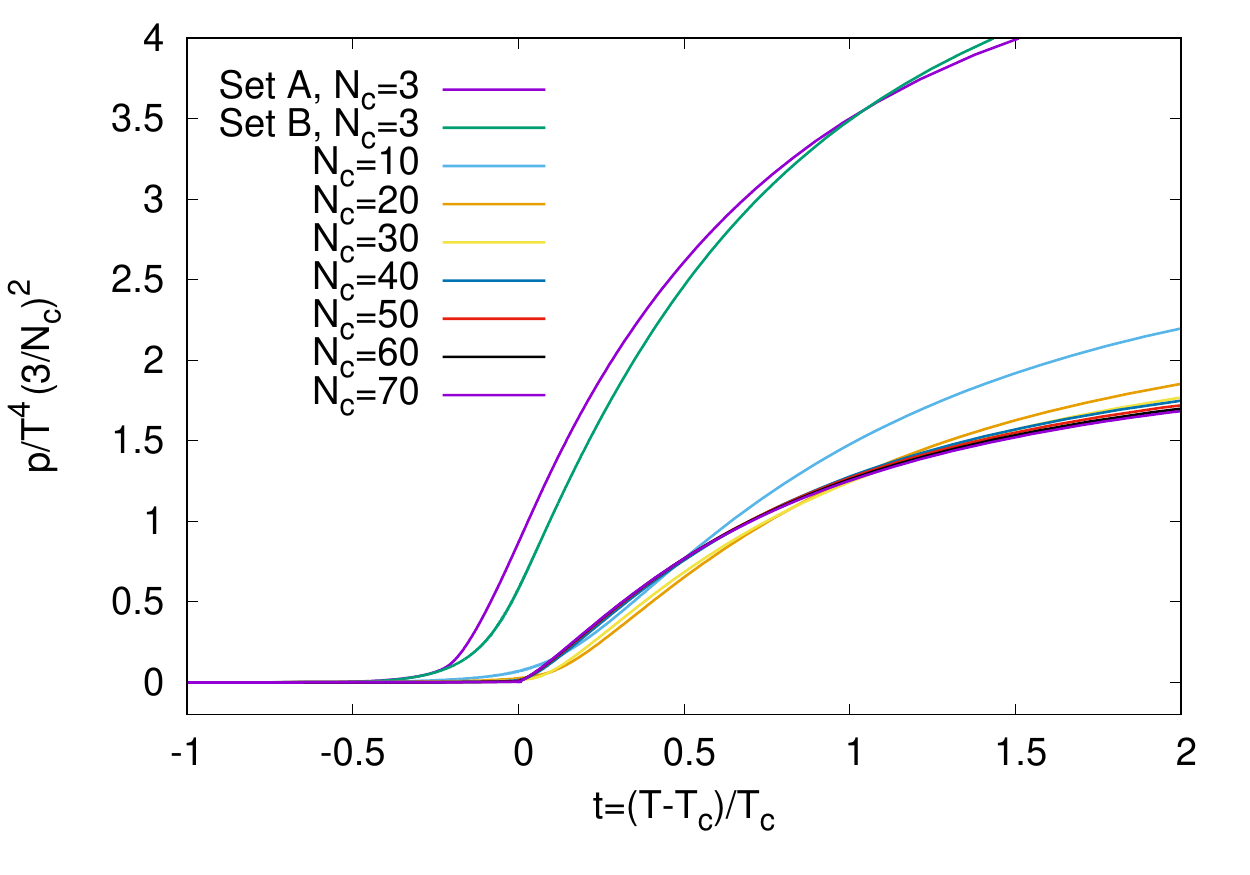}
    \caption{The normalized, rescaled pressure as a function of the reduced temperature.}
    \label{Fig:pressure_Fuea}
\end{figure}

\begin{figure}[htbp]
    \centering
    \includegraphics[width=0.45\textwidth]{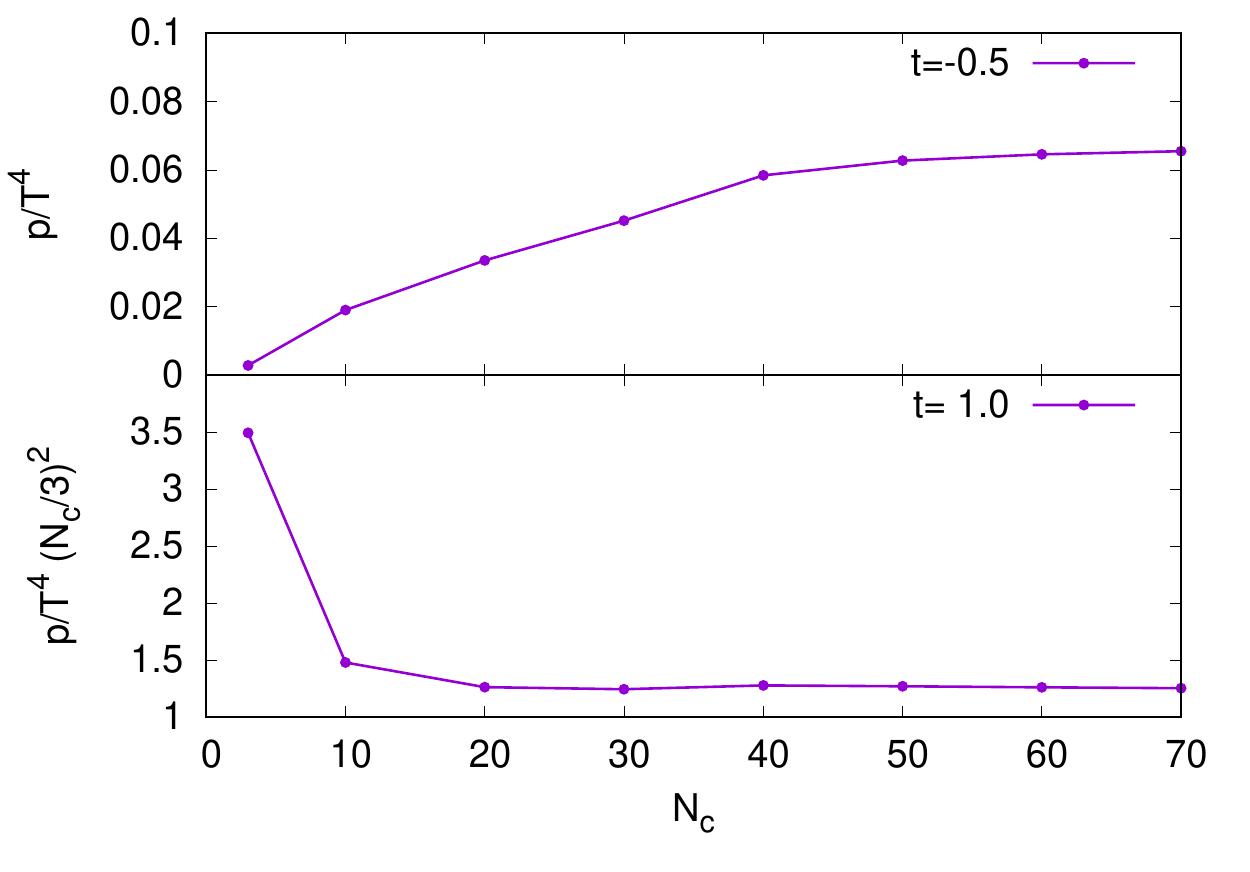}
    \caption{The normalized pressure as a function of $N_c$ for $t=-0.5$ (top) and $t=1$ (bottom).}
    \label{Fig:pressure_slices}
\end{figure}

Next, let us turn to the pressure, which is usually defined as
\begin{align}
    p(T, \mu_q) =& -\left(\Omega(T,\mu_q, \phi_{N/S}(T,\mu_q), \gamma(T,\mu_q))\nonumber \right.\\
    &\left.-\Omega(0,0, \phi_{N/S}(0,0), \gamma(0,0))\right),
\end{align}
where it is explicitly written that the grand potential depends both explicitly and implicitly---through the three order parameters, $\phi_N$, $\phi_S$ and $\gamma$---on $T$ and $\mu_q$. It turns out that if we use this definition for small temperatures the pressure has a leading linear $T$ dependence from 
\be U_\conf (T\ll T_c,\gamma) \propto T \ln H(\gamma=\gamma_0)=T \ln N_c^{N_c} .
\ee 
It is common to remove this linear part and use $U_\conf \propto T \ln \left(H/N_c^{N_c}\right)$ form in the Polyakov potential as in \cite{Roessner:2006xn, Sasaki:2012bi, Haas:2013qwp}. On the other hand, if we do this by redefining the potential in Eq.~\eqref{Eq:Grand_uea_Pol}, the pressure eventually becomes negative on some intervals, a feature that is physically not acceptable. This non-monotonic behavior of the pressure is a known problem for this kind of potentials, see e.g. \cite{Sasaki:2012bi}. However, the definition
\begin{align} \label{Eq:new_def_pressure}
     p(T, \mu_q) =& -\left(\Omega(T,\mu_q, \phi_{N/S}(T,\mu_q), \gamma(T,\mu_q))\nonumber \right.\\
    &\left.-\Omega(T,\mu_q, \phi_{N/S}(0,0), \gamma(0,0))\right)
\end{align}
solves both the problem of the linear temperature dependence and the negative pressure, since a nontrivial temperature dependent term is subtracted. Moreover, this term is independent of the order parameters, therefore the field equations---thus, the resulting phase structure---remain the same. 
The normalized rescaled pressure ($\frac{p}{T^4}(\frac{3}{N_c})^2$) as a function of the reduced temperature ($t=(T-T_c)/T_c$) is shown in Fig.~\ref{Fig:pressure_Fuea} for different $N_c$ values.  
It can be seen that the curves converge to the same curve for large $N_c$ above $T_c$, which shows the $N_c^2$ scaling of the pressure in the deconfined region. This behavior can also be seen in the bottom figure of Fig.~\ref{Fig:pressure_slices} at $t=1$. Here, in the top figure the $N_c^0$ scaling is realized in the confined region for sufficiently large $N_c$ at $t=-0.5$.

\subsection{The phase diagram and the fate of the critical end point(s)}

\begin{figure}[htbp]
    \centering
    \includegraphics[width=0.45\textwidth]{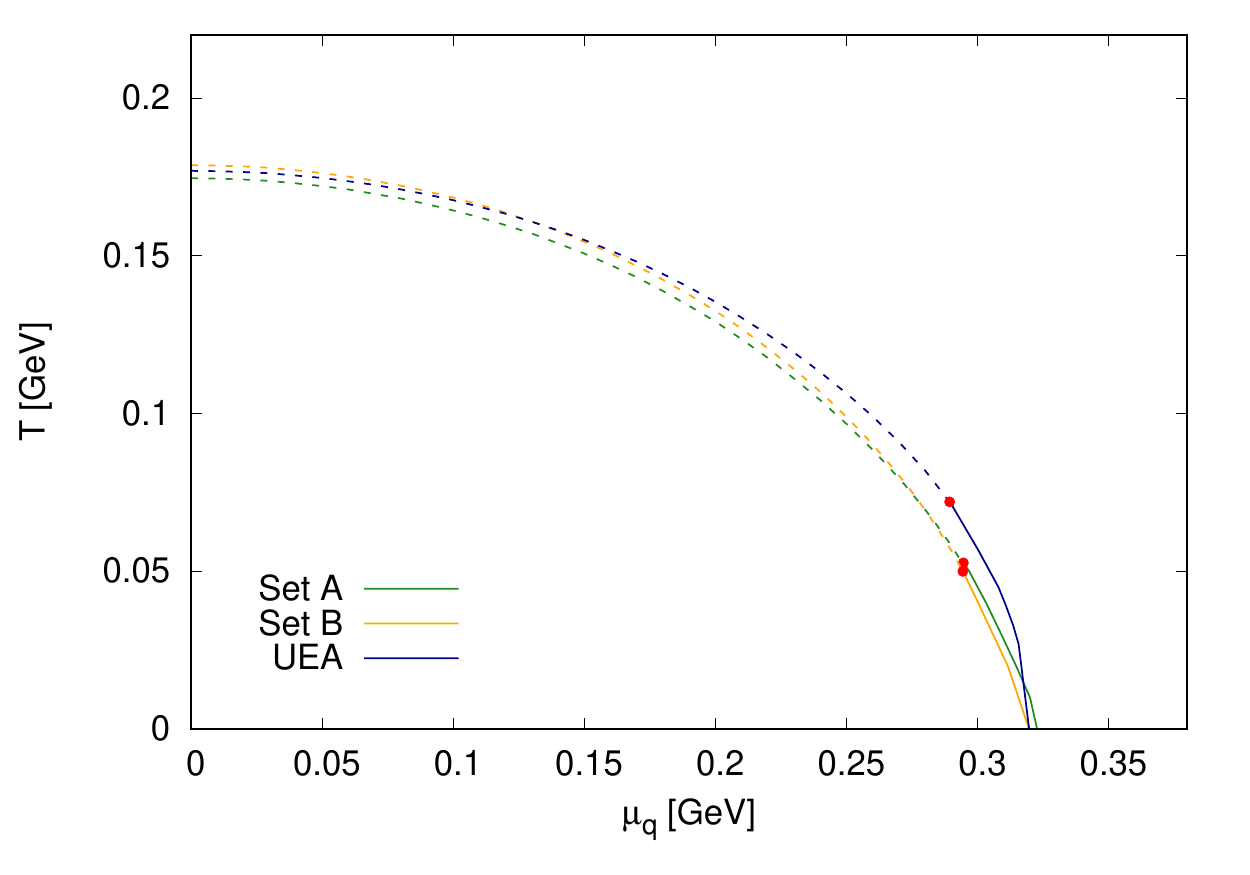}
    \caption{The phase diagram for three different cases at $N_c=3$, using set A and set B with Polyakov-potential from \cite{Kovacs:2016juc} and using set B with Polyakov-potential of Sec.~\ref{Ssec:Polyakov_Nc} (labeled with key 'UEA'). Dashed line refers to crossover, while solid line refers to first order phase transition}
    \label{fig:Fuea_pt}
\end{figure}
\begin{figure*}[ht!] 
    \centering
    \includegraphics[width=0.48\textwidth]{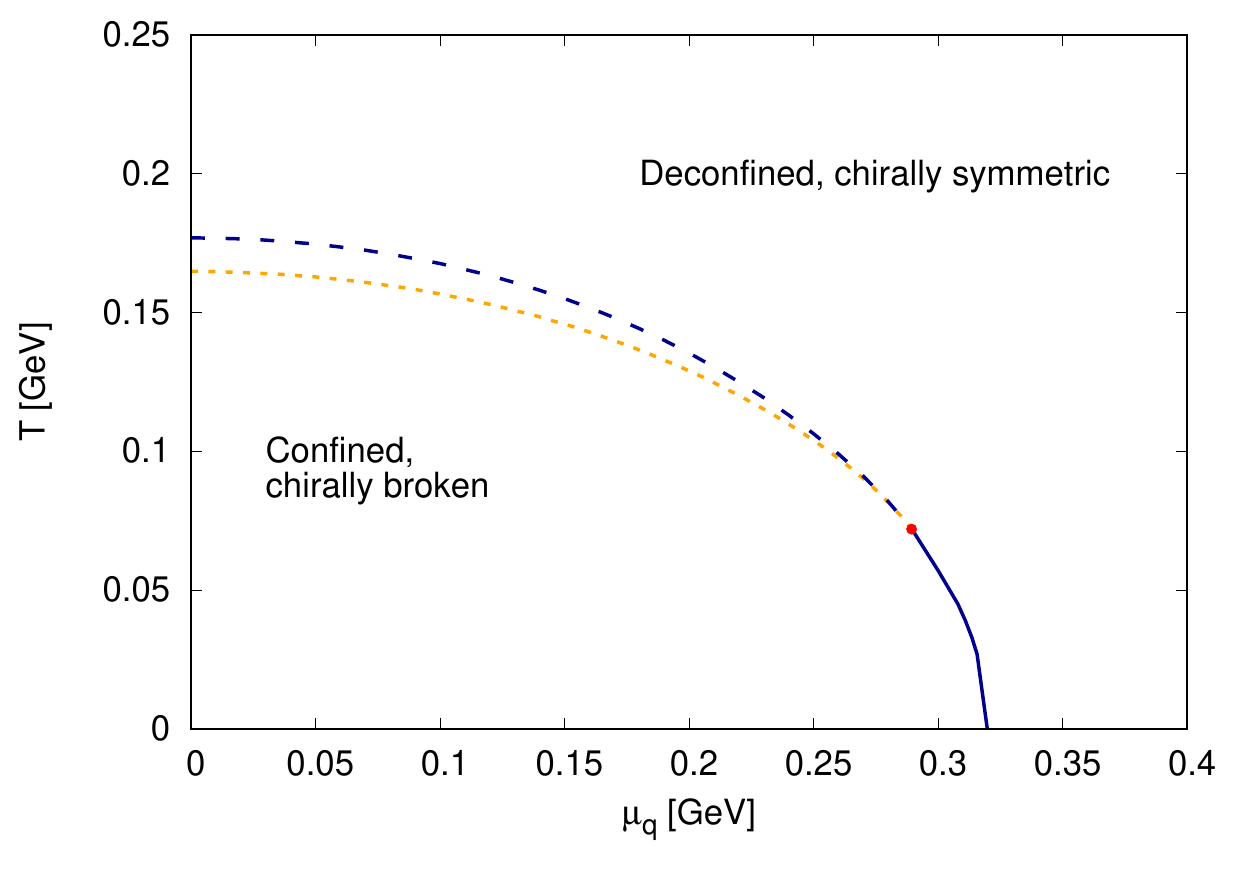}
    \includegraphics[width=0.48\textwidth]{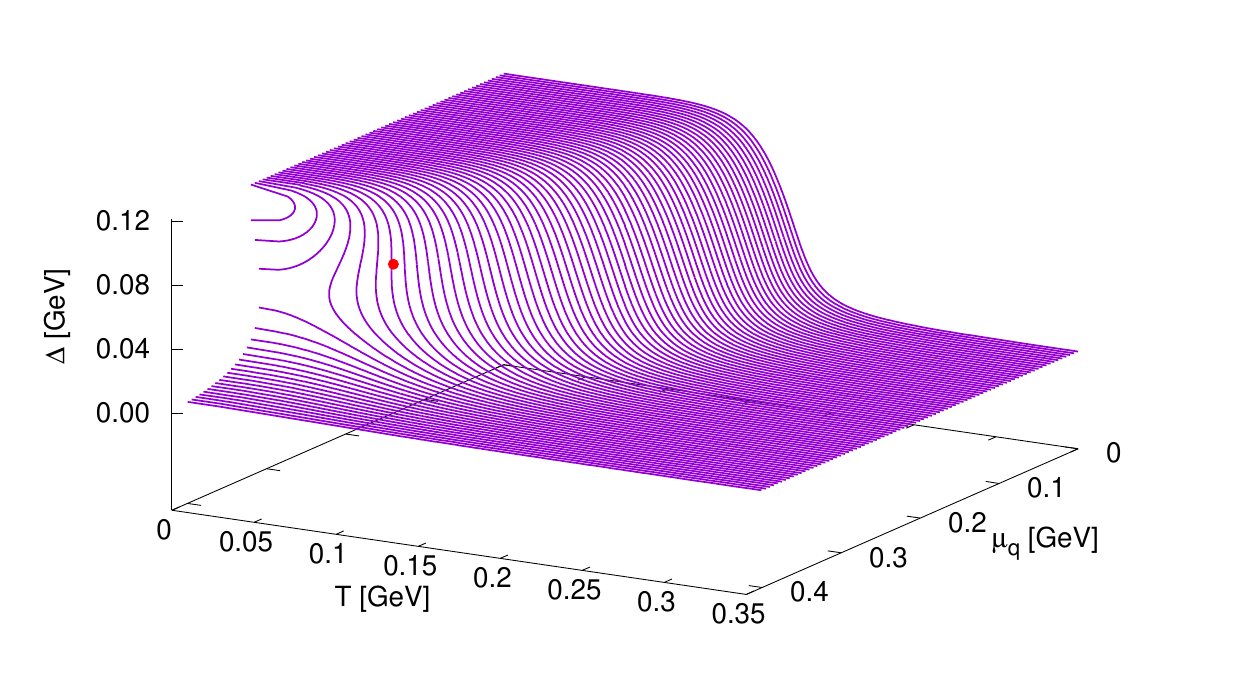}
    \includegraphics[width=0.48\textwidth]{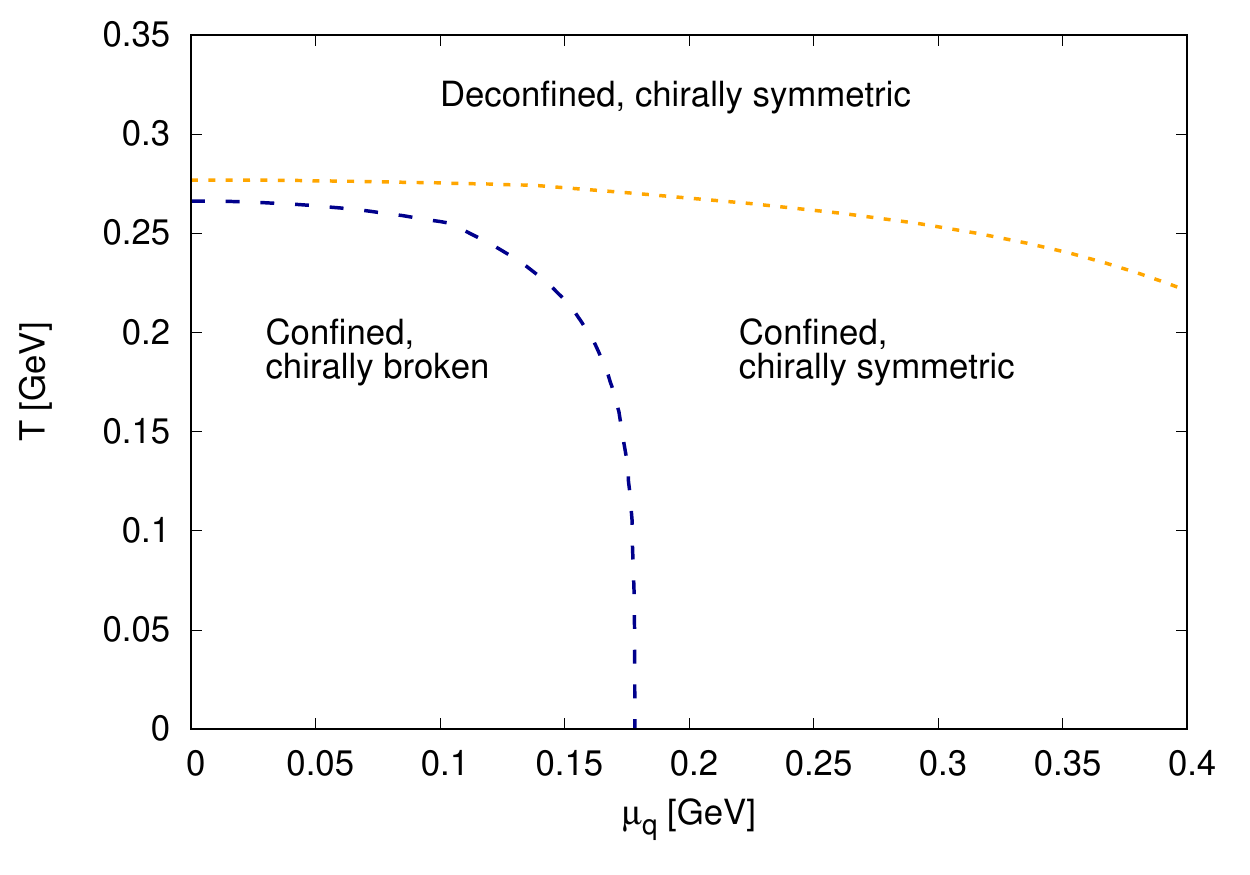}
    \includegraphics[width=0.48\textwidth]{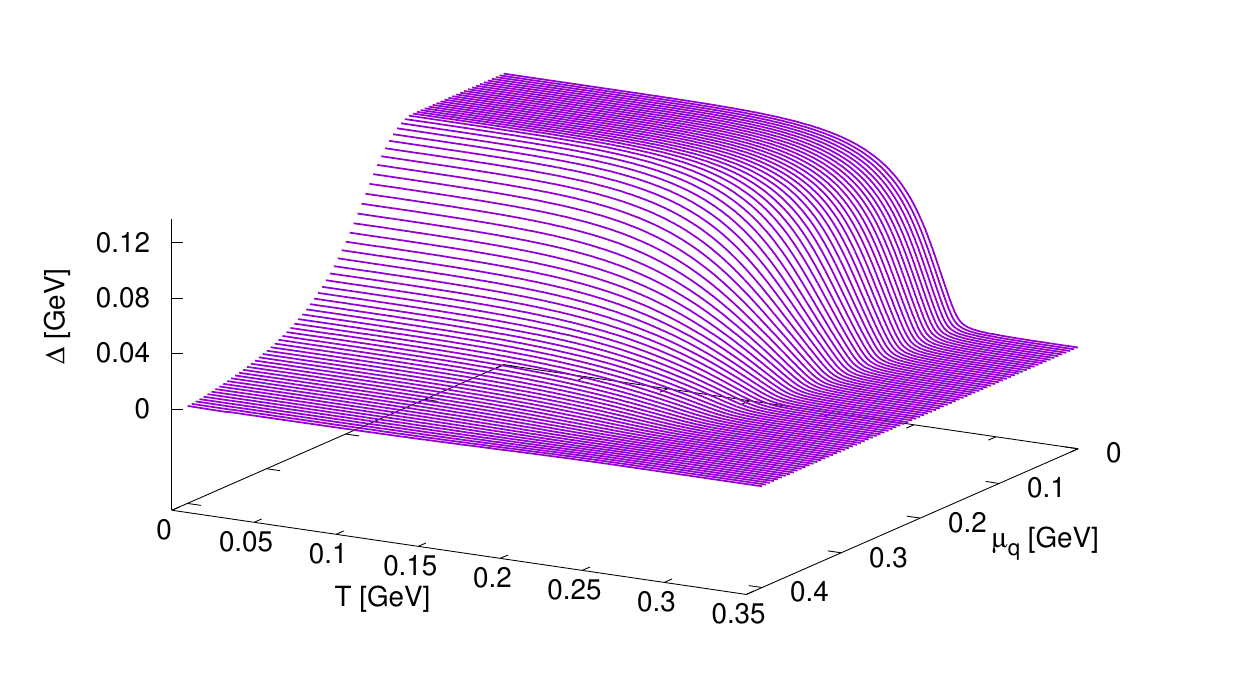}
    \includegraphics[width=0.48\textwidth]{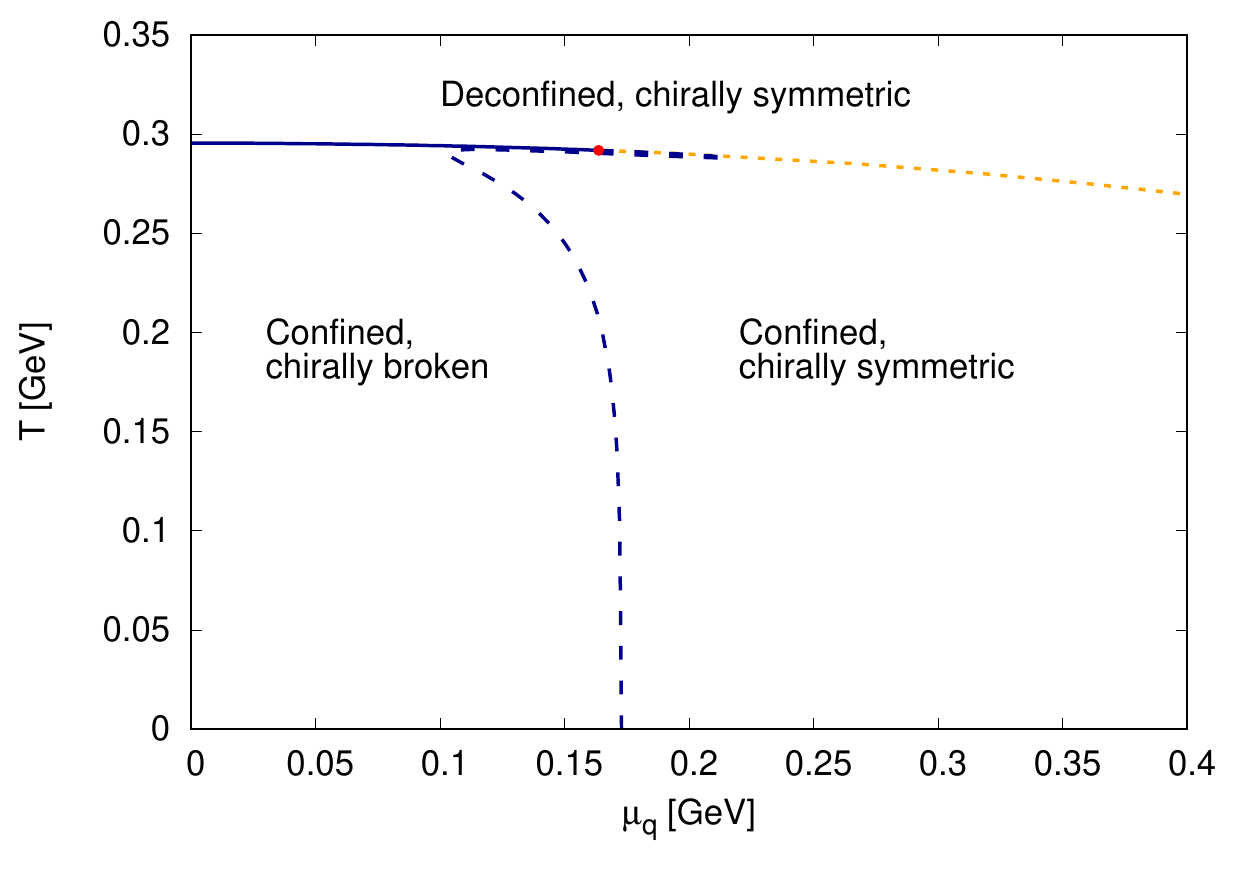}
    \includegraphics[width=0.48\textwidth]{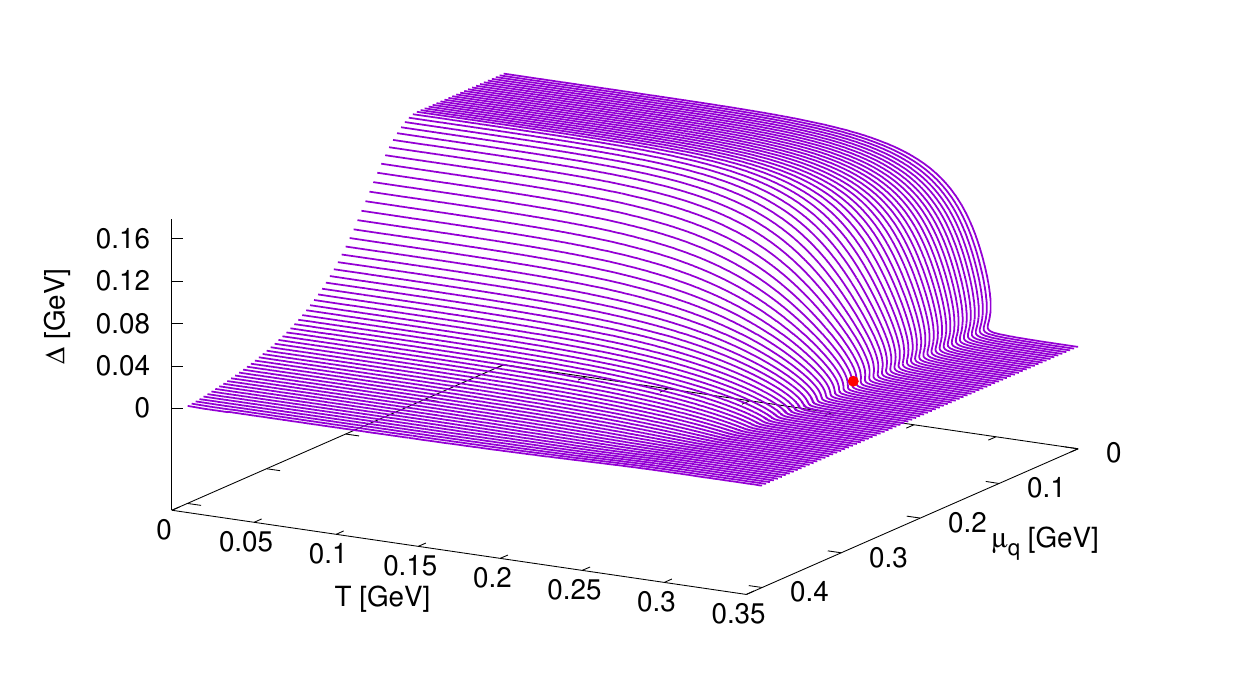}
    \caption{The phase diagram (left) for $N_c=3$ (top), $N_c=33$ (middle), $N_c=63$ (bottom), and their corresponding 3D plots for $\Delta$ (right). The dashed (solid) lines denote crossover (first order) type chiral phase transition, while the dotted line shows the deconfinement phase transition.}
    \label{fig:UEA_pt_Nc}
\end{figure*}
\begin{figure*}[ht!] 
    \centering
    \includegraphics[width=0.48\textwidth]{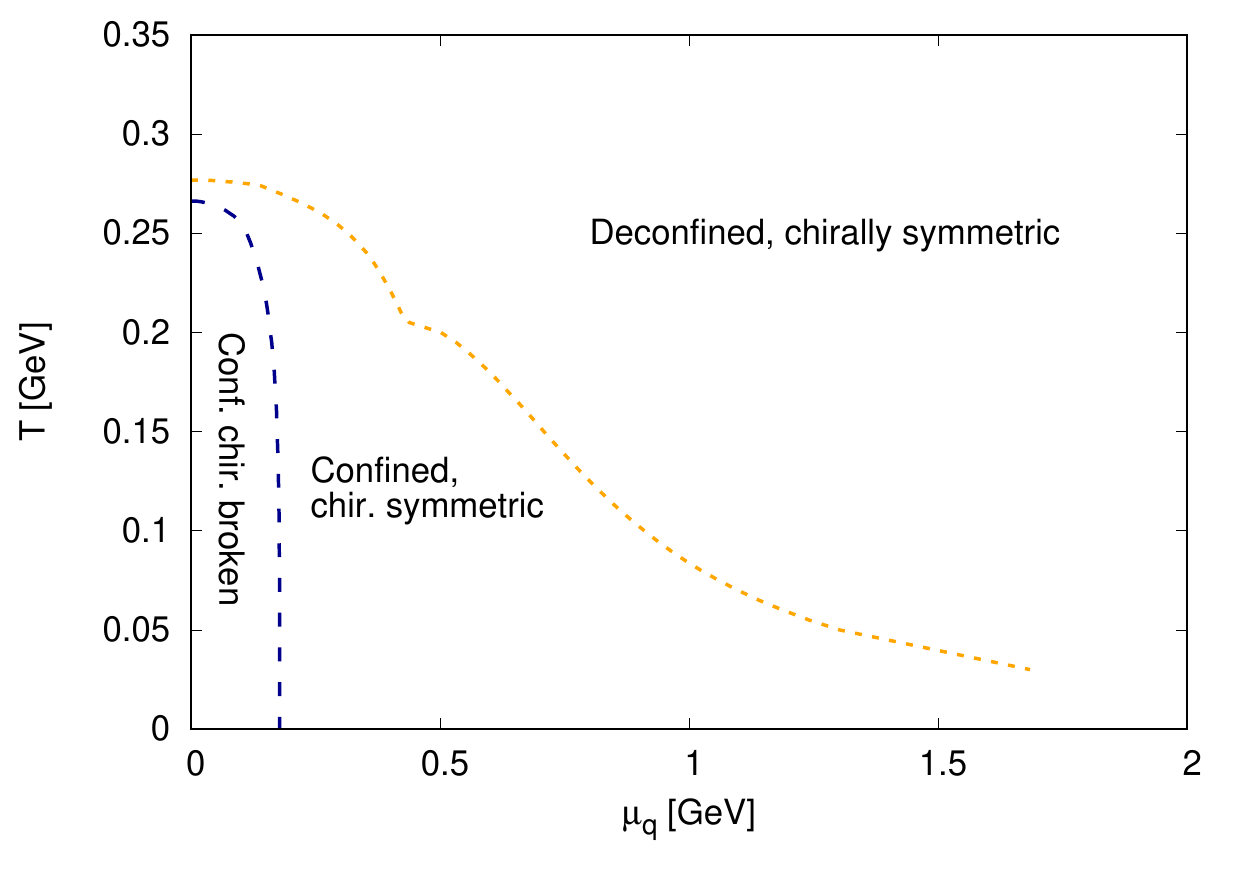}
    \includegraphics[width=0.48\textwidth]{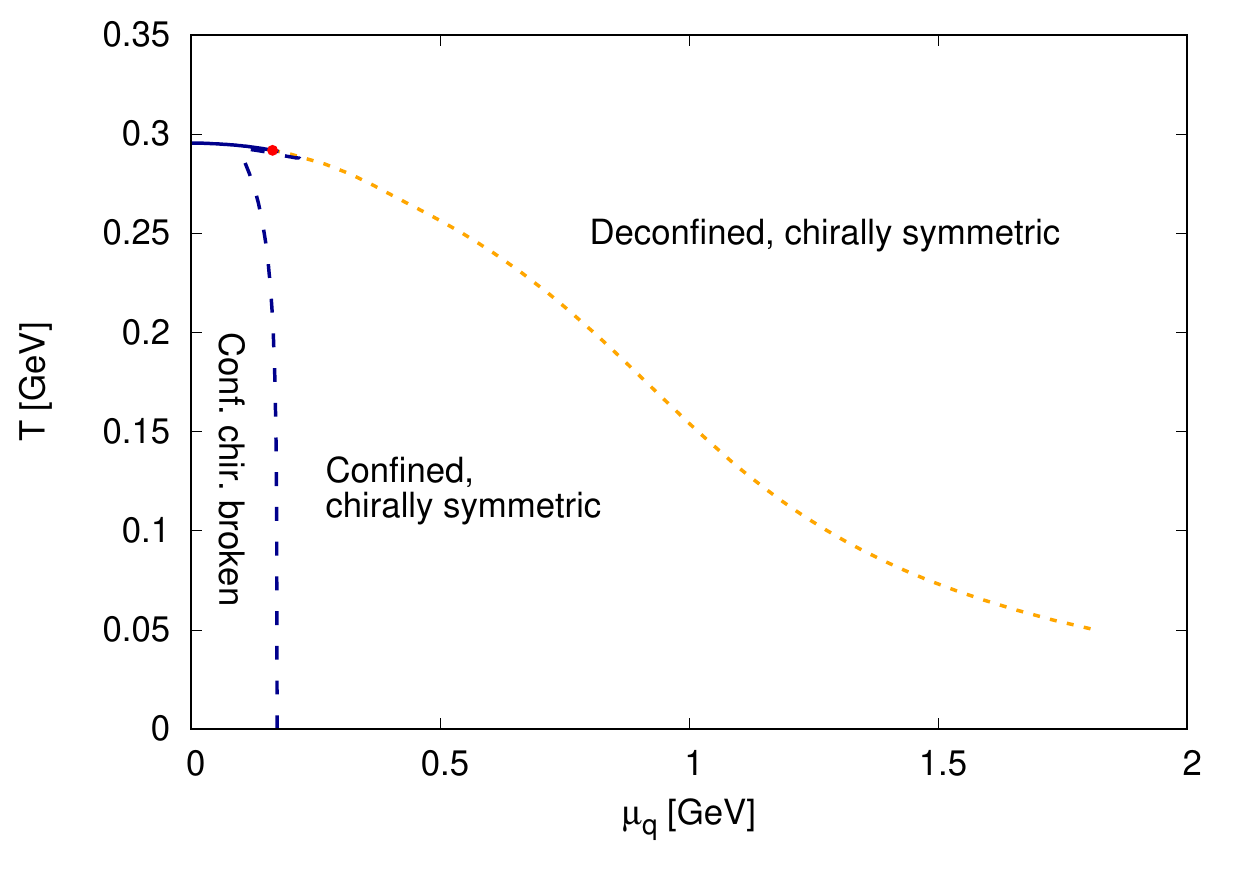}
    \caption{The phase diagram at $N_c=33$ and $63$ for a larger $\mu_q$ interval.}
    \label{fig:UEA_pt_Nc_Pol}
\end{figure*}
\begin{figure*}[ht!] 
    \centering
    \includegraphics[width=0.48\textwidth]{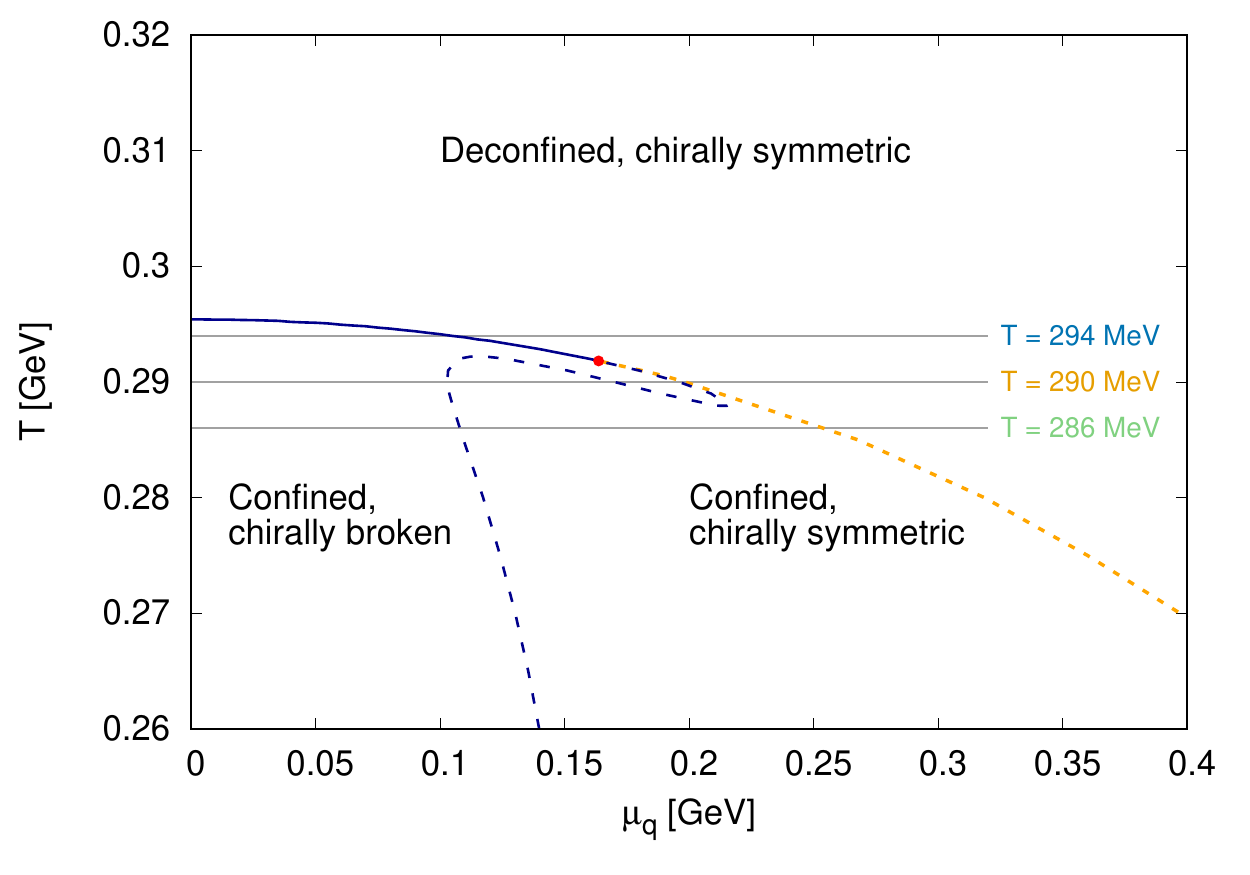}
    \includegraphics[width=0.48\textwidth]{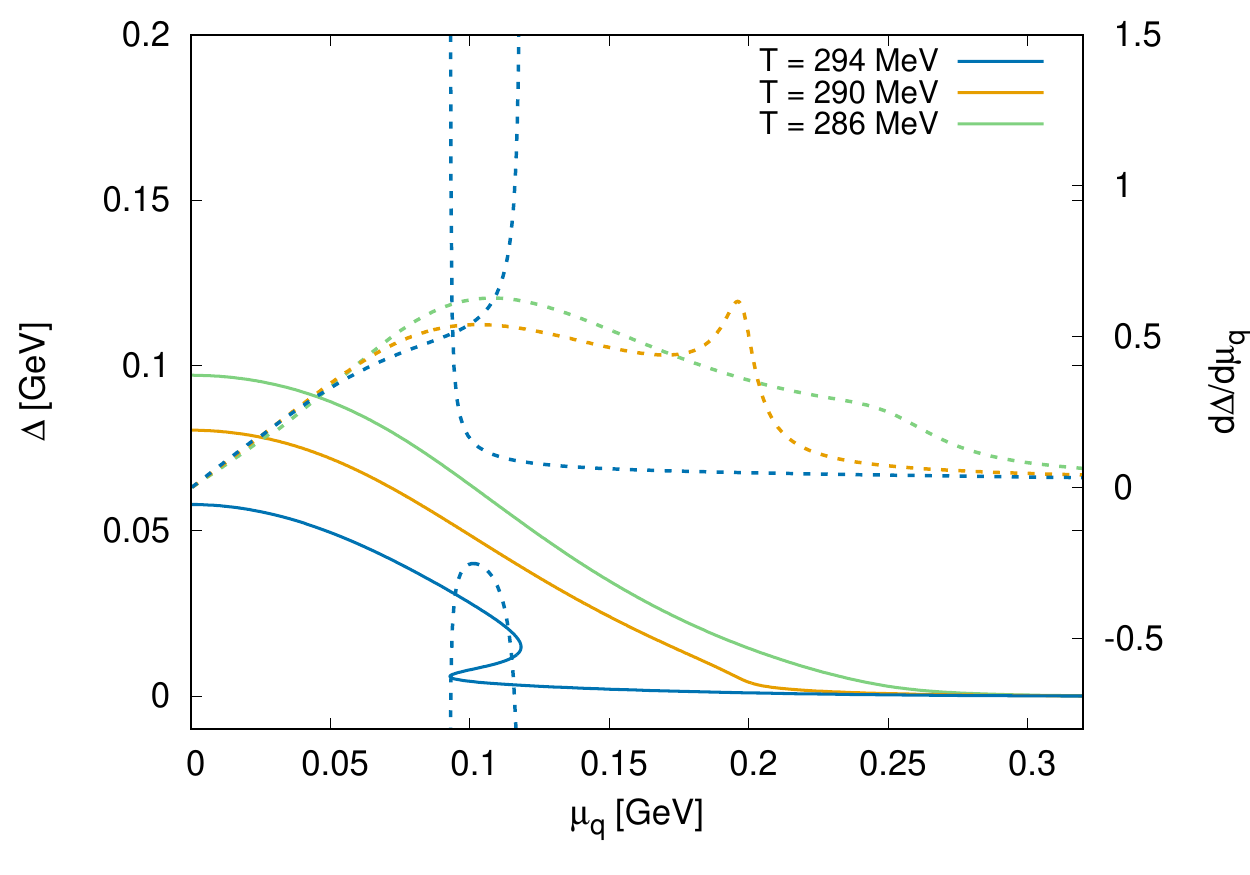}
    \caption{A closer look on the phase structure on the bottom left figure of Fig.~\ref{fig:UEA_pt_Nc} around the new CEP (left). The subtracted condensate $\Delta(T_0,\mu_q)$ (solid lines) and its $\mu_q$ derivative (dashed lines) calculated along the horizontal--$\mu_q$--direction (right).}
    \label{fig:struct_of_arm}
\end{figure*}
Now we turn to the investigation of the phase diagram on the $T$-$\mu_q$ plane, see Fig,~\ref{fig:Fuea_pt}. 
Three different setup are shown for $N_c=3$, two of them with the same Polyakov-loop potential used in \cite{Kovacs:2016juc} and with parameters of set A and set B of Tab~\ref{Tab:param}, while the third one with Polyakov-loop used in the uniform eigenvalue Ansatz (see Eq.~\eqref{Eq:UPol_uea}) and with set B. The dashed parts of the curves refer to crossover type phase transitions, while the solid ones refer to first order type phase transitions. The phase boundary lines are defined as the set of inflection points of the subtracted condensate defined as

\begin{equation}
\label{Eq:sub_cond}
    \Delta(T, \mu_q^{\text{fix}}) = \frac{(\phi_N - \frac{h_N}{h_S}\phi_S)|_{T, \mu_q^{\text{fix}}}}{(\phi_N - \frac{h_N}{h_S}\phi_S)|_{T=0,\mu_q^{\text{fix}}}} \text{ .}
\end{equation}
This quantity can be measured on the lattice \cite{Cheng:2007jq} and was already implemented in the PLeLSM in in \cite{Kovacs:2016juc}. In addition, this definition is applied when moving along the $T$ direction at a given fixed $\mu_q^{\text{fix}}$. Similarly, we also calculate this quantity along the $\mu_q$ direction for a given $T^{\text{fix}}$, for which the definition is the same with $T$ and $\mu_q$ interchanged. It is worth to note that location of inflection points of the subtracted condensate $\Delta$ and the nonstrange condensate $\phi_N$ are very close to each other. 

It can be noticed that there is a very small difference between curves made with the two different parameter sets and with the same Polyakov-loop potential, while the third curve using the UEA approximation is just a little above the other two. The large dots mark the critical end points on the different curves which are at $(\mu_q^\text{CEP},T^\text{CEP})=(294,50)$~MeV, $(295,53)$~MeV, and $(289,72)$ MeV from bottom to top. While the change in the value of $\mu_q^\text{CEP}$ is less than $5\%$, the change in the value of $T^\text{CEP}$ is more than $40\%$ for the two different Polyakov-loop potential.  

In Fig.~\ref{fig:UEA_pt_Nc} the phase boundaries are shown for the subtracted condensate (Eq.~\eqref{Eq:sub_cond}) and for the Polyakov-loop variables for different $N_c$ values. 
In this way we can distinguish different regions on the phase diagram, namely confined and chirally broken, confined and chirally symmetric, and deconfined and chirally symmetric. For $N_c=3$ there are basically two regions---beside a small region, which is deconfined, but chirally broken---, one which is confined and chirally broken, i.e. the normal baryonic matter and a deconfined, chirally symmetric quark-gluon plasma phase. The phase boundary is a crossover up to the large $\mu_q$ and small $T$ region, where the phase transition is of first order. The two regions are separated by a second order critical endpoint or CEP. With the increase of $N_c$, this picture changes very rapidly, the CEP disappears already for $N_c=4$, as it was already shown in Sec.~\ref{Ssec:res_zero_T}. For the illustrative value of $N_c=33$ there are only crossover type of transitions. In this case the confined and chirally broken phase shrink and a chirally symmetric but still confined phase appears. Quite interestingly, this phase can be interpreted as the quarkyonic phase of \cite{McLerran:2007qj}.
It should be noted here that when we define the phase boundary we either go along the $T$ direction starting from $T=0$ at some given ${\mu_q}_0$ or along the $\mu_q$ direction starting from $\mu_q=0$ at some given $T_0$. In case of crossover the location of inflection points of $\Delta(T,\mu_q)$ along the two different directions--$T$ or $\mu_q$--usually not coincide.

If we further increase $N_c$, at around $N_c=53$ a new CEP appears at $\mu_q=0$ along the $T$ axis, then this CEP starts to wander along the phase boundary towards larger $\mu_q$ and---compared to $T_c(\mu_q=0)$---smaller $T$ values. This can be seen on the bottom left figure of Fig.~\ref{fig:UEA_pt_Nc}. The right set of figures in Fig.~\ref{fig:UEA_pt_Nc} show the phase boundary surfaces of the $\Delta$ subtracted condensate--for the corresponding $N_c$ values. The phase diagram for $N_c=33$ and $63$ on a wider $\mu_q$ interval is displayed in Fig.~\ref{fig:UEA_pt_Nc_Pol} to show the high chemical potential behavior of the deconfinement phase transition. 
When the new CEP appears, there is a first order transition for high $T$, as a closer look to the case for $N_c=63$ in the left figure of Fig.~\ref{fig:struct_of_arm} shows.

Here the peculiarly shaped crossover line---defined via the inflection point along the $\mu_q$ direction---is also shown. Additionally, the right figure shows the subtracted condensate and its derivative along the horizontal lines of the left figure. There is a region in $T$, where $\Delta$ has multiple inflexion points (extrema of its derivative) which results in this unusual shape of the phase boundary. This means that, with increasing $\mu_q$ and after entering the chirally symmetric phase, the system would go back to the chirally broken region although the condensate is monotonically decreasing. This behavior highlights the limitation of defining the phase boundary univocally in case of a crossover. On the other hand, using the inflexion point is still the most common method to define a phase boundary both in the first order and the crossover region, thus being applicable in the whole phase diagram.

\begin{figure}[htbp]
    \centering\includegraphics[width=0.48\textwidth]{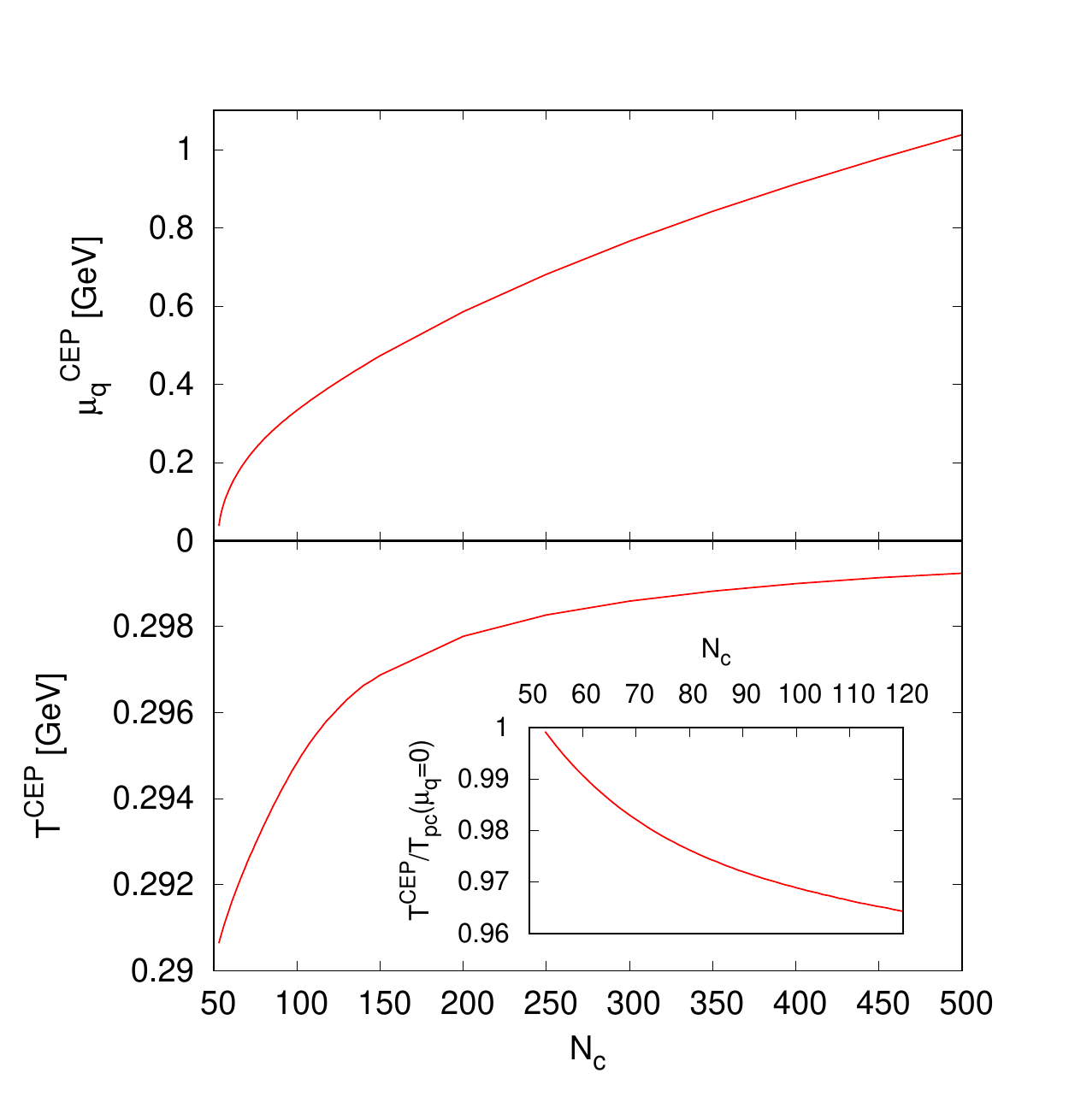}
    \caption{The $N_c$ dependence of $\mu_q^\text{CEP}$ (top) and $T^\text{CEP}$ (bottom). The inset shows $T^\text{CEP}/T_c(\mu_q=0)$.}
    \label{fig:CEP2_a}
\end{figure}

As already mentioned, the ``new'' large-$N_c$ induced CEP moves to higher quark chemical potentials, as can be seen in the top figure of Fig.~\ref{fig:CEP2_a}. 
The calculations were done again only at integer values of $N_c$ (the discrete points are connected for a better illustration). As the top figure of Fig.~\ref{fig:CEP2_a} shows, $\mu_q^\text{CEP}$ is increasing and no saturation can be seen. However, $T^\text{CEP}$ does saturate---bottom figure of Fig.~\ref{fig:CEP2_a}---for large $N_c$. In the inset of the bottom figure $T^\text{CEP}/T_c(\mu_q=0)$ is displayed, which shows that $T^\text{CEP}$ is decreasing less and less compared to the pseudocritical temperature at $\mu_q=0$ with increasing $N_c$. This behavior is also in agreement with the quarkyonic phase at large-$N_c$, according to which the first-order line becomes a horizontal line of the type  $T=T_c$, since the CEP moves toward an infinitely large chemical potential. Thus, our model realizes the expectations of quarkyonic matter.
\begin{figure}[ht!]
    \centering\includegraphics[width=0.45\textwidth]{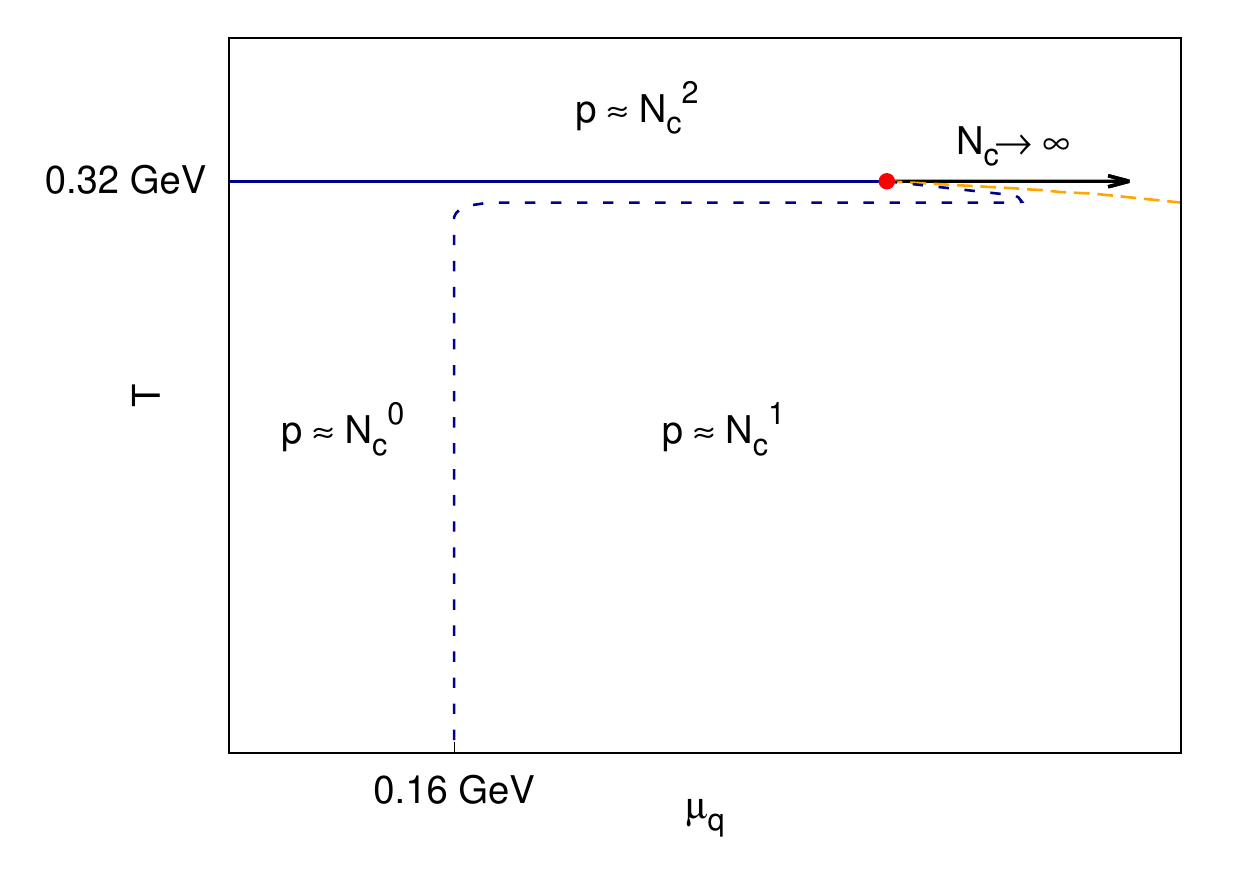}
    \caption{ The schematic phase diagram for large $N_c$ and the $N_c$ scaling of the pressure in the different phases.}
    \label{Fig:schematic}
\end{figure}
\begin{figure}[ht!]
    \centering\includegraphics[width=0.45\textwidth]{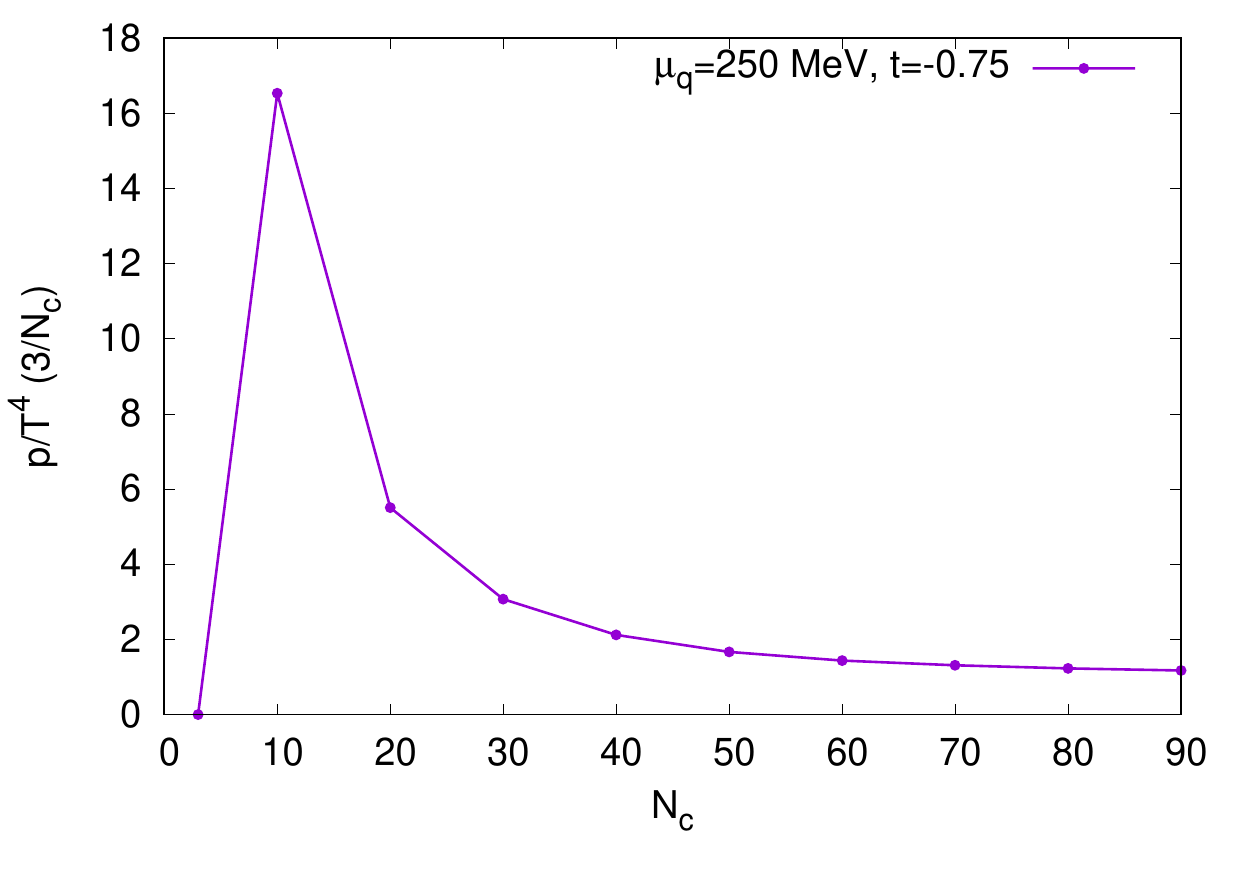}
    \caption{ The normalized, rescaled pressure as a function of $N_c$ for $\mu_q=250$~MeV and $t=-0.75$.}
    \label{Fig:p_quarkonia}
\end{figure}


The schematic phase diagram for large $N_c$ can be seen on Fig.~\ref{Fig:schematic}, where the $N_c$ scaling of the various phases are also marked. For small $T$ and $\mu_q$ one finds a meson dominated---confined and chirally broken---phase with $p\propto N_c^0$ pressure. It is separated by a first order chiral and deconfinement phase boundary around $T_c\approx 320$~MeV from the high temperature phase of quarks and gluons with $p\propto N_c^2$. The first order line ends in a second order critical endpoint that moves to higher chemical potential with the growing $N_c$. This CEP separates the first order and crossover boundaries both for the chiral and the deconfinement phase transitions. However, while the deconfinement boundary continues to larger chemical potentials and slowly approaching to smaller temperatures, the chiral boundary turns first to smaller $\mu_q$ then to small $T$ reaching the axis around $\mu_q^c\approx 160$~MeV and marks the start of a low temperature, higher chemical potential phase of confined but chirally restored matter. This region has a $p\propto N_c^1$ (see Fig.~\ref{Fig:p_quarkonia}) scaling, as expected for a quarkyonic phase. Note however, that this quarkyonic-like phase is separated from the deconfined phase with only either a very smooth crossover (large $\mu_q$) or a first-order boundary with nonzero Polyakov loop even in the low temperature side.

\section{Conclusions}
\label{sec:conclusion}

In this work, we investigated the large-$N_c$ behavior of the Polyakov loop extended linear sigma model (PLeLSM). When the parameters of the model are properly rescaled, the PLeLSM reproduces correctly the expected $N_c$ scaling of the physical quantities in the vacuum, such as the meson masses, decay widths, and the pion and kaon decay constants. 

Then, we concentrated on the phase diagram on the $\mu_q$-$T$ plane with increasing $N_c$ as well as on other relevant thermodynamic quantities, such as the $N_c$ scaling of the pressure. In particular, we have shown that the phase diagram at large-$N_c$ is substantially different from the physical case for $N_c=3$.

Along $T=0$---where the Polyakov loop decouples and the phase structure can be studied only with mesonic d.o.f. and their condensates---the chiral phase transition along the $\mu_q$ axis turns out to be a crossover already at $N_c=4$, which suggests that the critical end point disappears by then. 
This picture is confirmed when the whole phase diagram is studied and also the Polyakov loop---besides the meson condensates---plays an important role. Upon using the uniform eigenvalue Ansatz \cite{Dumitru:2012fw} (needed to reduce the number of d.o.f. and to cast the fermion determinant in a manageable form) as well as the Polyakov-loop potential of \cite{Lo:2021qkw},
we verified that the CEP actually disappears already at $N_c=4$ resulting in a crossover type transition on the whole $T$-$\mu_q$ plane. This is the phase diagram for intermediate values of $N_c$: a cross-over in all directions. In turn, this result implies that the CEP separating a cross-over on its left and first-order on its right is only a peculiarity of our natural world for $N_c=3$.

Next, when $N_c$ is sufficiently large ($N_c = 53$ within our model) the transition at $\mu_q=0$ eventually turns to a first order. This give rise to a new large-$N_c$ driven critical endpoint, which for increasing $N_c$  approaches a $N_c$-independent critical temperature $T_c$ of about 0.3 GeV and moves to higher chemical potential (eventually approaching infinity along the $\mu_q$ line).
Note, the fact that the critical temperature $T_c(\mu_q=0)$ (and similarly for each $\mu_q$) is large-$N_c$ independent, is due to the presence of the Polyakov loop and quark d.o.f., which resolve an apparent paradox emerging when only mesons are considered \cite{Heinz:2011xq}, which would---erroneously---imply that the critical temperature scales as $T_c^{1/2}$.

In the confined and chirally broken region at low $\mu_q$ and $T$, the pressure scales as $N_c^0$, while in the deconfined region ($T>T_c \sim0.3$~GeV) it scales as as $N_c^2$. This is in agreement with basic expectations, since the confined phase is dominated by mesons, while the deconfined phase is dominated by gluons for $N_c\to \infty$. The numerical value  $T_c \sim0.3$~GeV is also in agreement with the pure Yang-Mills results of \cite{Borsanyi:2012ve}.
Moreover, for large chemical potential $\mu_q>\mu_q^c \sim 0.16$~GeV and $T<T_c$ there is a confined and chirally symmetric phase, whose pressure is proportional to $N_c$: this is an explicit model realization of the quarkyonic phase at large-$N_c$. This is depicted in Fig.~\ref{Fig:schematic} that summarizes the phase-diagram in the large-$N_c$ limit. 

In the future, it is promising to study the restoration of dilatation invariance in the QGP phase diagram within the framework of the PLeLSM and neutron star matter \cite{McLerran:2018hbz,Marczenko:2022jhl,Fujimoto:2022ohj}.

\section*{Acknowledgments}

Gy. K. is thankful to Pok Man Lo for valuable discussions on the subject.
This research was supported by the Hungarian National Research, Development and Innovation Fund under Project No. FK 131982. F. G. acknowledges support from the Polish National Science Centre (NCN) through the OPUS project no 2019/33/B/ST2/00613. 

\appendix

\section{Terms in $g^{\pm}_f$ up to $N_c\geq 8$}
\label{App:g_f_Nc_8}

In this Appendix we list more contributions to $g^+$ that was introduced in Section~\ref{sec:polyakov_Nc}.
One can continue with the $n=3$ and $N_c-3$ phases and then the $n=4$ and $N_c-4$ phases, where the calculations becomes more and more complicated, because it is harder to express everything with the Polyakov loop parameters of Eq.~\eqref{Eq:phi_N_def}. Accordingly, one finds for the $n=3$ part of $g^+$:
\begin{align} \nonumber
g^+\Big|_{n=3}=&\frac{1}{6} \Bigg(\sum_{\substack{a,b,c }} e^{-i( q_a+q_b+q_c)} \\ \nonumber
&\quad- 3\sum_{\substack{a,b }} e^{-i( q_a+2q_b)} +2 \sum_a e^{-i3q_a}\Bigg) e^{-3\beta E_f^+}\\
=&\frac{1}{6}\left( N_c^3 \bar\Phi^3 -3N_c^2 \bar\Phi \bar\Phi_2 +2 N_c \bar\Phi_3\right) e^{-3\beta E_f^+}.
\end{align}
Similarly, for the $n=N_c-3$ part of $g^+$, one has:
\begin{align} \nonumber
g^+ & \Big|_{n=N_c-3}=\\ \nonumber
&=\frac{1}{6} \Bigg(\sum_{\substack{a,b,c }} e^{i( q_a+q_b+q_c)} \\ \nonumber
&\qquad - 3\sum_{\substack{a,b }} e^{i( q_a+2q_b)} +2 \sum_a e^{i3q_a}\Bigg) e^{-(N_c-3)\beta E_f^+}\\
&=\frac{1}{6}\left( N_c^3 \Phi^3 -3N_c^2 \Phi \Phi_2 +2 N_c \Phi_3\right) e^{-(N_c-3)\beta E_f^+}.
\end{align}
Continuing this procedure the contribution to the color determinant from the terms with $n=4$ phases is
\begin{align} \nonumber
g^+\Big|_{n=4} = \frac{1}{24}&\Big( N_c^4 \bar \Phi^4 -6 N_c^3 {\bar \Phi}^2 \bar \Phi_2 + 8 N_c^2\bar \Phi \bar \Phi_3 \\&+ 3 N_c^2 {\bar \Phi_2}^2 -6 N_c \bar \Phi_4 \Big)e^{-4\beta E_f^+} \text{ ,}
\end{align}
while for $n=N_c-4$ (changing to $n'=N_c-n=4$) 
\begin{align} \nonumber
g^+\Big|_{n=N_c-4} = \frac{1}{24}&\Big( N_c^4 \Phi^4 -6 N_c^3 {\Phi}^2 \Phi_2 + 8 N_c^2\Phi \Phi_3 \\&+ 3 N_c^2 {\Phi_2}^2 -6 N_c \Phi_4 \Big)e^{-(N_c-4)\beta E_f^+}.
\end{align}
Consequently, for $N_c\geq 8$ one gets
\begin{align} \label{Eq:general_det_dagger_upto4}
g^+=& 1+e^{-N_c\beta E_f^+} \nonumber \\
& +N_c\left[\bar{\Phi} e^{-\beta E_f^+}+\Phi e^{-(N_c-1)\beta E_f^+}\right] \nonumber \\
&+ \frac{1}{2}\left( N_c^2 \bar \Phi^2 - N_c \bar \Phi_2 \right) e^{-2\beta E_f^+} \nonumber \\
&+ \frac{1}{2}\left( N_c^2 \Phi^2 - N_c \Phi_2 \right) e^{-(N_c-2)\beta E_f^+} \nonumber \\
&+\frac{1}{6}\left( N_c^3 \bar\Phi^3 -3N_c^2 \bar\Phi \bar\Phi_2 +2 N_c \bar\Phi_3\right) e^{-3\beta E_f^+} \nonumber \\
&+\frac{1}{6}\left( N_c^3 \Phi^3 -3N_c^2 \Phi \Phi_2 +2 N_c \Phi_3\right) e^{-(N_c-3)\beta E_f^+} \nonumber \\
&+\frac{1}{24}\Big( N_c^4 \bar \Phi^4 -6 N_c^3 {\bar \Phi}^2 \bar \Phi_2 + 8 N_c^2\bar \Phi \bar \Phi_3 \nonumber \\
&\qquad\qquad\qquad+ 3 N_c^2 {\bar \Phi_2}^2 -6 N_c \bar \Phi_4 \Big)e^{-4\beta E_f^+} \nonumber \\
&+\frac{1}{24}\Big( N_c^4 \Phi^4 -6 N_c^3 {\Phi}^2 \Phi_2 + 8 N_c^2\Phi \Phi_3 \nonumber \\
&\qquad\qquad\qquad+ 3 N_c^2 {\Phi_2}^2 -6 N_c \Phi_4 \Big)e^{-(N_c-4)\beta E_f^+} \nonumber \\
&+[\text{terms with 5 to Nc-5 phases}] \text{ ,}
\end{align}
and the expression for $g^-$ differs again from $g^+$ only in changing $\bar{\Phi}\leftrightarrow \Phi$ and $-\mu_q \rightarrow +\mu_q$.

\section{$\Phi_n\approx \Phi^n$ approximation}
\label{App:phi_n_phi_n}

The uniform eigenvalue Ansatz is a very useful tool to study the large $N_c$ limit of the PLeLSM. However, in this approximation it is assumed that the Polyakov loop variables are purely real, which is not the case for the $\mu_q>0$ calculations already at $N_c=3$. In general, one needs to allow $\Phi_n \neq \bar \Phi_n$. Therefore we investigated another approximation, where the original fermion determinant is expressed in terms of the Polyakov-loop parameters as in Eq.~\eqref{Eq:general_det_dagger_upto4} but the emerging higher powers are simply replaced by using the $\Phi_n=\Phi^n$ Ansatz. This is clearly not exact for $N_c\geq4$, but it can be shown that the error $\left| \Phi_n-\Phi^n \right|$ is limited, while in the coefficients of $g_f^\pm$ it is suppressed by an extra factor of $N_c^m$ with $1\leq m \leq n-1$. In the $\Phi_n\approx \Phi^n$ approximation $g_f^+$ in Eq.~\eqref{Eq:general_det_dagger_upto4} can be rewritten as\footnote{Note that due to the sum this numeric calculation takes a longer time and the factorials lead to very large number therefore only $N_c\leq170$ is applicable to avoid the overflow.}
\begin{align}\label{Eq:Omega_repld} \nonumber
g^+=& 1+e^{-N_c\beta E_f^+} \\ &+ \sum_{n=1}^N \binom{N_c}{n} \left(\bar \Phi^n e^{-n\beta E_f^+} + \Phi^n e^{-(N_c-n)\beta E_f^+} \right)
\end{align}
while $g^-$ can be written again from $g^+$ only by changing $\bar{\Phi}\leftrightarrow \Phi$ and $-\mu_q \rightarrow +\mu_q$. We note that the same form can be obtained by replacing each $e^{\pm iq_\alpha}$ phase factors in the sum of phases in Eq.~\eqref{Eq:gf+Nc1} to its averaged value $\sum_{i=1}^{N_c} e^{\pm iq_i}/N_c=\frac{1}{N_c}\Tr_c L^\pm =\Phi^\pm$, where $L^-=L$, $L^+=L^\dagger$, $\Phi^-=\Phi$ and $\Phi^+=\bar \Phi$. For the Polyakov-loop potential in this approximation we use the same form employed in \cite{Kovacs:2016juc} with an additional artificial $N_c^2$ scaling 
\begin{align} \label{Eq:UPolPaw} \nonumber
U_\Pol =&T^4 \left(\frac{N_c}{3}\right)^2\Big[-\frac{1}{2} a(T) \Phi \bar \Phi \\&+ b(T) \ln \left( 1 - 6\Phi \bar\Phi + 4 (\Phi +\bar \Phi^3) - 3 (\Phi\bar\Phi)^2 \right)\Big] \text{ ,}
\end{align}
with the coefficients 
\be
a(T)= a_0 + a_1 \left(\frac{T_0}{T}\right) +a_2\left(\frac{T_0}{T} \right)^2, \quad b(T) =b_3 \left( \frac{T_0}{T} \right)^3 \text{ ,}
\ee 
where the values of the constants are $a_0=3.51$, $a_1=-2.47$, $a_2=15.22$ and $b_3=-1.75$.
This $N_c$ scaling is expected to reproduce the $N_c$ dependence of the pressure in the deconfined phase. This is also supported by the leading $N_c^2$ scaling of the Polyakov potential in Eq.~\eqref{Eq:UPol_uea}.

\begin{figure}[htbp]
    \centering\includegraphics[width=0.48\textwidth]{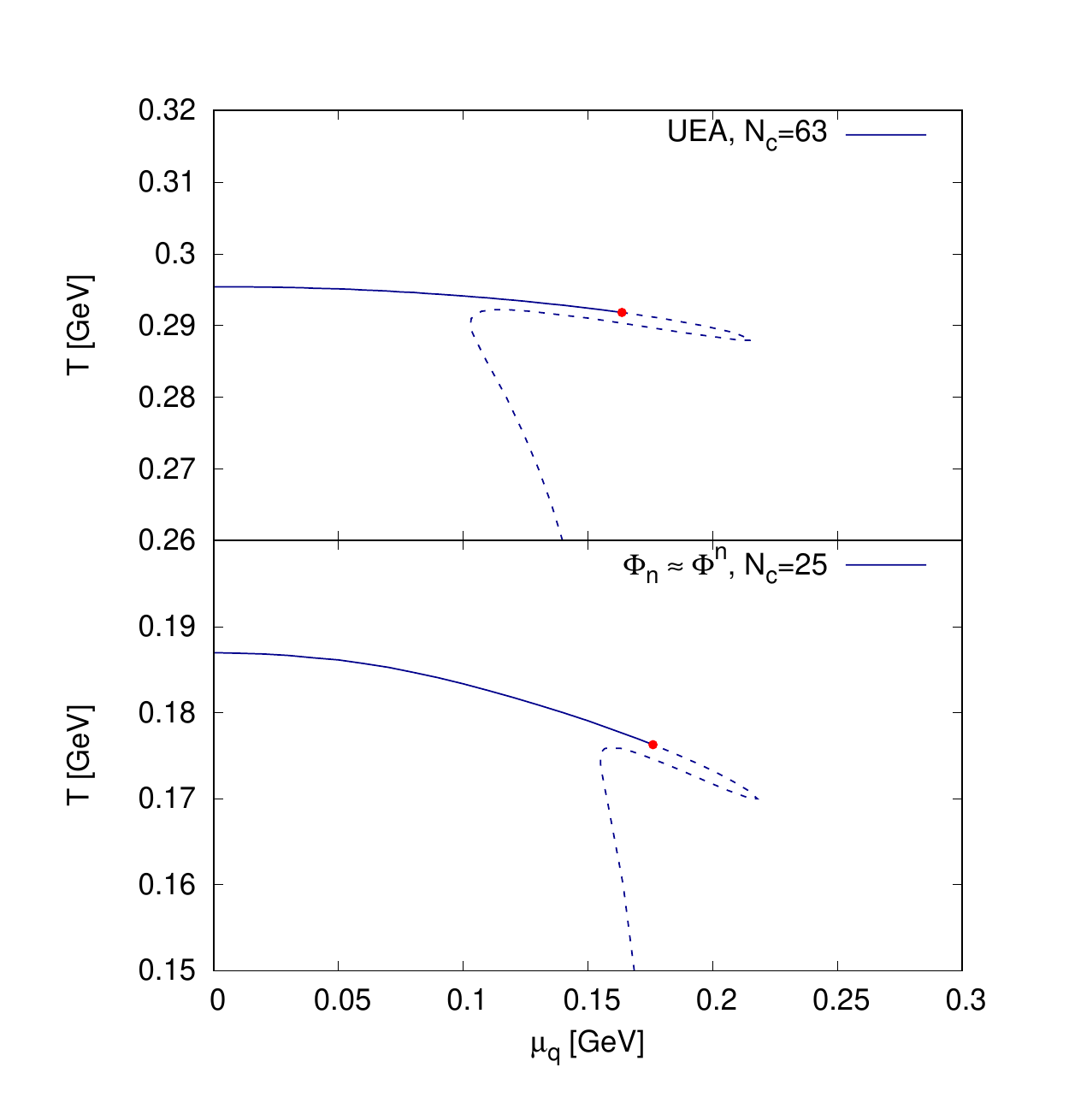}
    \caption{The phase diagram using the uniform eigenvalue Ansatz and the $\Phi_n\approx\Phi^n$ approximation. Note the different scales on the vertical axis.}
    \label{fig:comp_UEA_nn}
\end{figure}

At $N_c=3$ the $\Phi_n\approx \Phi^n$ approximation reduces to the model employed in \cite{Kovacs:2016juc}, whose  corresponding  phase diagram is shown in Fig.~\ref{fig:Fuea_pt}. 
With growing $N_c$ the transition become crossover for the entire phase boundary\footnote{The disappearance of the CEP can be seen from the $T=0$ behavior presented in Fig.~\ref{fig:phiN_muB_zeroT} and therefore independent of the used Polyakov sector.} and a second critical end point separating a first order transition emerges for low $\mu_q$ just as in the case of the uniform eigenvalue Ansatz, thus, showing the same qualitative behavior. On the other hand, the appearance of the second CEP takes place already for $N_c=15$ due to the artificial scaling of the Polyakov potential being stronger for intermediate $N_c$ values than the scaling of the potential in Eq.~\eqref{Eq:UPol_uea}. For a comparison Fig.~\ref{fig:comp_UEA_nn} shows the phase diagram for both of our approximations at an $N_c=63$ and $N_c=25$ for the UEA and the $\Phi_n\approx \Phi^n$ approximation. The $N_c$ values were chosen to be well above, but not far from the appearance of the second critical end point, 
and the qualitative similarity is not influenced by the actual choice of $N_c$. Note that the range of the temperature axis differs significantly since the critical temperature at vanishing chemical potential---and therefore the first order line---saturates at a different value in the large $N_c$ limit.

\FloatBarrier

\bibliography{ELSM_large_Nc}

\end{document}